\documentclass[twocolumn]{aastex62}

\usepackage{graphicx,amsmath}
\usepackage{amssymb}
\usepackage{nccmath}
\usepackage{upgreek}
\usepackage{subfigure}
\usepackage{color}
\usepackage{bm}
\usepackage{soul}
\usepackage{CJK}
\usepackage{ulem}
\hypersetup{linkcolor=blue, citecolor=blue}

\setcitestyle{notesep={ }}
\newcommand{\Msol}{\hbox{\thinspace $M_{\odot}$}}

\newcommand{\htwo}{H$_2$}

\submitjournal{ApJ on August 31, 2020}

\shorttitle{The Fundamental Formation Relation}
\shortauthors{Dou et al.}

\begin{document}

\title{{\bf \Large From haloes to galaxies - II. The fundamental relations in star formation and quenching}}

\correspondingauthor{Yingjie Peng}
\email{yjpeng@pku.edu.cn}

\author[0000-0002-6961-6378]{Jing Dou}
\affiliation{Kavli Institute for Astronomy and Astrophysics, Peking University, 5 Yiheyuan Road, Beijing 100871, China}
\affiliation{Department of Astronomy, School of Physics, Peking University, 5 Yiheyuan Road, Beijing 100871, China}

\author{Yingjie Peng}
\affiliation{Kavli Institute for Astronomy and Astrophysics, Peking University, 5 Yiheyuan Road, Beijing 100871, China}

\author[0000-0002-7093-7355]{Alvio Renzini}
\affiliation{INAF - Osservatorio Astronomico di Padova, Vicolo dell'Osservatorio 5, I-35122 Padova, Italy}

\author[0000-0001-6947-5846]{Luis C. Ho}
\affil{Kavli Institute for Astronomy and Astrophysics, Peking University, 5 Yiheyuan Road, Beijing 100871, China}
\affil{Department of Astronomy, School of Physics, Peking University, 5 Yiheyuan Road, Beijing 100871, China}

\author[0000-0002-4803-2381]{Filippo Mannucci}
\affiliation{Istituto Nazionale di Astrofisica, Osservatorio Astrofisico di Arcetri, Largo Enrico Fermi 5, I-50125 Firenze, Italy}

\author[0000-0002-3331-9590]{Emanuele Daddi}
\affiliation{AIM, CEA, CNRS, Universit\'{e} Paris-Saclay, Universit\'{e} Paris Diderot, Sorbonne Paris Cit\'{e}, F-91191 Gif-sur-Yvette, France}

\author[0000-0003-0007-2197]{Yu Gao}
\affiliation{Purple Mountain Observatory $\&$ Key Laboratory for Radio Astronomy, Chinese Academy of Sciences, 10 Yuanhua Road, Nanjing 210033, PR China}
\affiliation{Department of Astronomy, Xiamen University, Xiamen, Fujian 361005, China}

\author[0000-0002-4985-3819]{Roberto Maiolino}
\affiliation{Cavendish Laboratory, University of Cambridge, 19 J. J. Thomson Avenue, Cambridge CB3 0HE, UK}
\affiliation{Kavli Institute for Cosmology, University of Cambridge, Madingley Road, Cambridge CB3 0HA, UK}

\author[0000-0001-6469-1582]{Chengpeng Zhang}
\affiliation{Kavli Institute for Astronomy and Astrophysics, Peking University, 5 Yiheyuan Road, Beijing 100871, China}
\affiliation{Department of Astronomy, School of Physics, Peking University, 5 Yiheyuan Road, Beijing 100871, China}

\author[0000-0002-3890-3729]{Qiusheng Gu}
\affil{School of Astronomy and Space Science, Nanjing University, Nanjing 210093, China.}
\affil{Key Laboratory of Modern Astronomy and Astrophysics (Nanjing University), Ministry of Education, Nanjing 210093, China.}

\author[0000-0003-3010-7661]{Di Li}
\affiliation{CAS Key Laboratory of FAST, National Astronomical Observatories, Chinese Academy of Sciences, Beijing 100012, China}
\affiliation{School of Astronomy and Space Science, University of Chinese Academy of Sciences, Beijing 100049, China}

\author[0000-0002-6423-3597]{Simon J. Lilly}
\affiliation{Department of Physics, ETH Zurich, Wolfgang-Pauli-Strasse 27, CH-8093 Zurich, Switzerland}

\author[0000-0003-3564-6437]{Feng Yuan}
\affiliation{Key Laboratory for Research in Galaxies and Cosmology, Shanghai Astronomical Observatory, Chinese Academy of Sciences, 80 Nandan Road, Shanghai 200030, China}

\begin{abstract}

\noindent  Star formation and quenching are two of the most important processes in galaxy formation and evolution. We explore in the local Universe the interrelationships among key integrated galaxy properties, including stellar mass $M_*$, star formation rate (SFR), specific SFR (sSFR), molecular gas mass $M_{\rm H_2}$, star formation efficiency (SFE) of the molecular gas and molecular gas to stellar mass ratio $\mu$. We aim to identify the most fundamental scaling relations among these key galaxy properties and their interrelationships. We show the integrated $M_{\rm H_2}$-SFR, SFR-$M_*$ and $M_{\rm H_2}$-$M_*$ relation can be simply transformed from the $\mu$-sSFR, SFE-$\mu$ and SFE-sSFR relation, respectively. The transformation, in principle, can increase or decrease the scatter of each relation. Interestingly, we find the latter three relations all have significantly smaller scatter than the former three corresponding relations. We show the probability to achieve the observed small scatter by accident is extremely close to zero. This suggests that the smaller scatters of the latter three relations are driven by a more universal physical connection among these quantities. We then show the large scatters in the former relations are due to their systematic dependence on other galaxy properties, and on star formation and quenching process. We propose the sSFR-$\mu$-SFE relation as the Fundamental Formation Relation (FFR), which governs the star formation and quenching process, and provides a simple framework to study galaxy evolution. Other scaling relations, including integrated Kennicutt-Schmidt law, star-forming main sequence and molecular gas main sequence, can all be derived from the FFR.

\end{abstract}


%-----------------------------------------------------------------------------------------------------------------------------
\section{Introduction} \label{sec:intro}

The assembly of the galaxy population across cosmic time can be mainly described by the star formation history \citep[e.g.,][for a review]{1996ApJ...460L...1L,2014ARA&A..52..415M}, quenching history \citep[e.g.,][]{2010ApJ...721..193P,2016MNRAS.460L..45R}, merging history \citep[e.g.,][for a review]{1993MNRAS.264..201K,2014ARA&A..52..291C}, chemical enrichment history \citep[e.g.,][for a review]{2019A&ARv..27....3M} and angular momentum history \citep[e.g.,][]{1969ApJ...155..393P,1984ApJ...286...38W,1998MNRAS.295..319M,2000ApJ...538..477N,2020MNRAS.495L..42R,2020MNRAS.491L..51P}. Stars are commonly believed to form in cold dense molecular clouds. Star formation rate (SFR) can be associated to the total available cold molecular gas mass ($M_{\rm H_2}$) in the interstellar medium (ISM) of the galaxy via star formation efficiency (SFE) as SFR = SFE $\times M_{\rm H_2}$. The SFE, or equivalently the gas depletion timescale $\tau$ ($\tau$ = 1/SFE), describes how efficiently the galaxy can convert the available cold gas into stars.

Many observations have shown that the SFR is tightly correlated with the gas content, especially the molecular gas. This fundamental scaling relation is often referred as the star formation law or Kennicutt-Schmidt (KS) law \citep{1959ApJ...129..243S,1998ApJ...498..541K}, and can be parameterized as $\Sigma_{\rm SFR} \propto \Sigma_{\rm H_2}^N $, where $\Sigma_{\rm SFR}$ and $\Sigma_{\rm H_2}$ are the star formation rate and molecular gas surface densities. The value of $N$ depends on different galaxy samples, different methods in measuring the SFR, different physical scales and different \htwo\ tracers. For example, when the surface densities are averaged over the entire galaxy, $N$ is found to be $\sim$ 1.4 for CO-based \htwo\ gas, while on kpc or sub-kpc scales, $N$ is $\sim$ 1 or lower for the same gas tracer \citep{2008AJ....136.2846B,2013AJ....146...19L,2017ApJ...846..159B,2018ApJ...863L..21K,2019MNRAS.488.1926D,2019ApJ...884L..33L,2020MNRAS.496.4606M}. When the dense molecular gas (traced by HCN or HCO+) is used, the star formation law is linear \citep{2004ApJ...606..271G}.

Although the KS star formation law was discovered about six decades ago, the physical origin of this tight relation is still debated. In theories, two scenarios have been proposed to explain the observations \citep{2012ARA&A..50..531K}. The bottom-up picture describes that the star formation is mainly driven by localized processes within giant molecular clouds \citep{2005ApJ...630..250K,2011ApJ...729..133M}. The top-down scenario assumes that the star formation is primarily determined by global, large-scale dynamical processes in galaxies, such as disk instabilities \citep{1997ApJ...481..703S}. However, there are recent results showing that the molecular KS relation holds also on sub-galactic scales \citep{2019ApJ...884L..33L,2019MNRAS.488.1926D,2020MNRAS.496.4606M,2020MNRAS.492.6027E,2020MNRAS.493L..39E}, which hence seem contrary to the top-down scenario.

In observations, the star formation law has been explored in increasing details by expanding the parameter space, i.e. to study the multiple interrelationships among SFR, $M_{\rm H_2}$, SFE (or $\tau$ in the equivalent), sSFR, molecular gas to stellar mass ratio $\mu$, stellar mass $M_*$, etc. For instance, $\tau$ of the molecular gas was first found to be about constant, 1 $\sim$ 2 Gyr in normal spiral galaxies \citep{2008AJ....136.2846B,2011AJ....142...37S,2013AJ....146...19L}, while subsequent studies have shown that the molecular $\tau$ in fact depends on various galaxy properties, in particular on sSFR (\citealp{2010ApJ...714L.118D,2010MNRAS.407.2091G,Genzel:2015fq,2013AJ....146...19L,2011MNRAS.415...61S,2013ApJ...778....2S,2016MNRAS.462.1749S,2014A&A...562A..30S,2014ApJ...793...19S,Huang:2014ko,2014MNRAS.445.2599B,2016NatAs...1E...3P,2016ApJ...833..112S,2016ApJ...820...83S,2017ApJ...837..150S}; \citealp[for a review]{2010Natur.463..781T,2013ApJ...768...74T,2018ApJ...853..179T,2020arXiv200306245T}; \citealp{2017A&A...604A..53C,2019A&A...622A.105F,2019ApJ...878...83W,Aravena_2019,2019ApJ...884..177Y,2020MNRAS.492L...6P}).

In line with this effort, several key scaling relations have been identified. The first one is the star-forming main sequence \citep{2004MNRAS.351.1151B,2007ApJ...670..156D,2007A&A...468...33E,2011A&A...533A.119E,2007ApJ...660L..43N,2007ApJS..173..315S,2008ApJ...688..770F,2010ApJ...721..193P,2012ApJ...754L..29W,2014ApJ...795..104W,2014ApJS..214...15S,2015ApJ...800L..10R,2015ApJ...801L..29R,2015A&A...575A..74S,2019MNRAS.483.3213P,2019MNRAS.490.5285P}, which describes the tight relation between SFR and stellar mass of the star-forming galaxies. The second one is the molecular gas main sequence \citep{2018ApJ...852...74B,2018ApJ...853..179T,2019ApJ...884L..33L,2020MNRAS.492.2651B}, which describes the correlation between molecular gas mass and stellar mass. The third one is the so-called extended KS relation, which describes the correlation between SFR and the combination of gas mass and stellar mass \citep{1985ApJ...295L...5D,1994ApJ...430..163D,2011ApJ...733...87S,2018ApJ...853..149S}. There are also increasing interests and many efforts that have been made to study the combined 3D relation defined by $M_*$, SFR and $M_{\rm H_2}$ \citep{2019ApJ...884L..33L,2020MNRAS.496.4606M,2020MNRAS.492.6027E,2020MNRAS.493L..39E}.

In this work, we focus on analyzing the underlying interrelationships among these key scaling relations, and paying particular attention to their scatters. We aim to identify the most important or universal scaling relations connecting different key galaxy properties with the smallest scatters, which can in return give insight into the underlying physics.

Since the star formation is observed to be more directly correlated with molecular gas \citep{1988ApJ...334..613S,2007ApJ...671..303B,2010ApJ...714L.118D,2010MNRAS.407.2091G}, in this work we first focus on molecular gas only and will include atomic gas in our analysis in a future study. Galaxy mergers can have a complicated effect on the star formation of the galaxies. It can enhance or suppress star formation, depending on the gas content of the merging galaxies and also the phase of the merger. The major merger rate is observed to increase with redshift \citep{Cibinel_2019,Ferreira_2020}, even up to z $\sim$ 6 \citep{Duncan_2019}, and the galaxy major merger fraction is only a few percent in the local Universe \citep{2017A&A...608A...9V,Duncan_2019}. On average, mergers are expected to have very small impact on the star formation and quenching in the local galaxy population. In this paper, we hence do not consider the effect of mergers in studying the scaling relations in star formation and quenching in the local Universe. However, the effect of mergers should be taken into account when moving to high redshift.

Throughout this work we assume the following cosmological parameters: $\Omega_m=0.3, \ \Omega_\Lambda=0.7, \ H_0=70\,\rm {km\,s^{-1} Mpc^{-1}}$.

%-----------------------------------------------------------------------------------------------------------------------------
\section{Sample} \label{sec:sample}

The molecular gas sample used in this paper is the Extended CO Legacy Database for the GALEX Arecibo SDSS Survey (xCOLD GASS, \citet{Saintonge:2017iz}). It contains 532 nearby galaxies with CO (1-0) observations over a total of $\sim$ 950 hours observing time using the IRAM 30m single-dish telescope. It is a compilation of two large CO programs. The original COLD GASS sample was selected randomly from the parent sample of SDSS spectroscopic survey \citep{Abazajian:2009ef} within the ALFALFA footprints \citep{Haynes:2011en}. It includes 366 galaxies with $M_*>10^{10}\Msol$ within the redshift range of $0.025<z<0.05$ \citep{2011MNRAS.415...32S}. The second large program, COLD GASS-low survey, extended the sample to a lower stellar mass range ($10^{9}\Msol<M_*<10^{10}\Msol$) at lower redshift ($0.01<z<0.02$). The lower redshift range was to ensure that the CO (1-0) line can be detected for the galaxies with low stellar mass. The angular sizes of the low-mass galaxies are also smaller and most of their CO flux are within the beam of the IRAM 30m telescope.

The CO observation and data reduction are described in \citet{Saintonge:2017iz} in detail. The general observation strategy was to observe a galaxy until either the CO (1-0) line was detected with S/N $>$ 5 or the RMS noise was low enough to put a stringent upper limit on the gas fraction, which was set to be $M_{\rm H_2}/M_* = 1.5\%$ for COLD GASS and 2.5$\%$ for COLD GASS-low. Only targets with FLAGCO = 1, i.e. the CO (1-0) line is detected and S/N $>$ 5, are used in our work. The selection effects are discussed in detail in Appendix A.

The beam size of IRAM telescope at 3 mm is 22''. For most galaxies, their flux can be recovered with a single pointing of the telescope. However, an aperture correction was still applied to all measured CO (1-0) line fluxes to account for the large angular size of some galaxies. The detailed method is described in \citet{2012ApJ...758...73S}. The median aperture correction for the whole xCOLD GASS sample is 1.17, close to unity. The aperture corrected CO (1-0) fluxes are then used to derive the total molecular gas mass.

We use the recommended value for the CO-to-\htwo\ conversion factor from xCOLD GASS, which was calculated using the calibration function in \citet{Accurso:2017kf}. It is metallicity-dependent and has a second order dependence on the offset from the star-forming main sequence. We have also tested alternative conversion factors, including constant $\alpha_{\rm CO}$ or metallicity-dependent only $\alpha_{\rm CO}$ \citep[e.g.,][]{2012ApJ...746...69G}. The results are very similar to these presented in the paper, hence our results are not sensitive to the choice of conversion factor.
% \begin{fleqn}
% \begin{equation}
% \begin{split}
%  \label{eq1}
% \rm log\ \alpha_{CO}= & \ 14.752\ - \ 1.623[12 \ + \ \rm log(O/H)] \\
%  &+ \ 0.062 \  \rm log \ \Delta(MS),
% \end{split}
% \end{equation}
% \end{fleqn}
%where 12 + log (O/H) is the gas-phase metallicity, $\Delta$(MS) is the distance offset from the star-forming main sequence.
This calibration is independent of any assumption on the molecular gas depletion timescale, and hence is ideal for studying gas and star formation scaling relations.

It should be noted that xCOLD GASS is a mass-selected sample, and it has a flat distribution in stellar mass, which is different from the mass distribution of the parent SDSS sample (i.e. stellar mass function with a negative slope at the low-mass end). This “mass bias” is corrected by a statistical weight \citep{2010MNRAS.403..683C}. For each xCOLD GASS galaxy, the weight is calculated as the ratio between the number of galaxies from the expected stellar mass function derived from the volume-limited sample and the actual number of objects in the xCOLD GASS sample at a given stellar mass bin.

Stellar masses of the xCOLD GASS galaxies are taken from the SDSS DR7 MPA-JHU catalog, derived from fits to the photometry \citep{2007ApJS..173..267S}. SFRs are calculated using the combination of MIR and UV from WISE and GALEX survey databases, as described in \citet{Janowiecki:2017bb}. Using a different SFR (e.g., SFR based on WISE+SDSS+GALEX from \citet{2016ApJS..227....2S}) produces very similar results. Also our primary goal is to compare the relative differences between different scaling relations, not to determine their absolute slope and normalization, hence systematic uncertainties in SFRs are less critical. All stellar masses and SFRs are converted to a Chabrier initial mass function \citep{2003PASP..115..763C}. The final sample used in our analysis contains 330 galaxies with reliable CO (1-0), SFR and stellar mass measurements. This is smaller than the original sample size of xCOLD GASS that contains 532 galaxies, mainly due to the CO non-detections of many galaxies below the main-sequence as discussed in Appendix A.

We also extract structural parameters for each galaxy. The r-band effective radius (R$_{50}$) is obtained from the SDSS DR7 official database. The mass weighted bulge-to-total ratios (B/T) are taken from \citet{2014ApJS..210....3M}, in which each galaxy is fitted with a pure exponential disk and a de Vaucouleurs bulge (Sersic index $n_b$ = 4). The value of B/T is defined as $M_{bulge}$ / ($M_{bulge}$ + $M_{disk}$). We also checked that the total stellar mass from ($M_{bulge}$ + $M_{disk}$) agrees well with the mass directly derived from the SED-fitting method \citep{2004MNRAS.351.1151B}, with a small scatter of only 0.1 dex on average.

It should be noted that unlike many previous similar studies where only star-forming galaxies are selected, in our analysis we do not differentiate between star-forming and passive galaxies. We include all galaxies from star-bursting ones and down to those with the lowest observable SFR and \htwo\ gas mass, though fully quenched galaxies are probably not included as most of them are not detected in CO. One of our main goals is to test whether the star formation physics is the same or not in star-forming galaxies and galaxies in the process of being quenched.

%-----------------------------------------------------------------------------------------------------------------------------
\section{$M_{\rm H_2}$-SFR v.s. $\mu$-sSFR relations}

\begin{figure*}[htbp]
    \begin{center}
       \includegraphics[width=180mm]{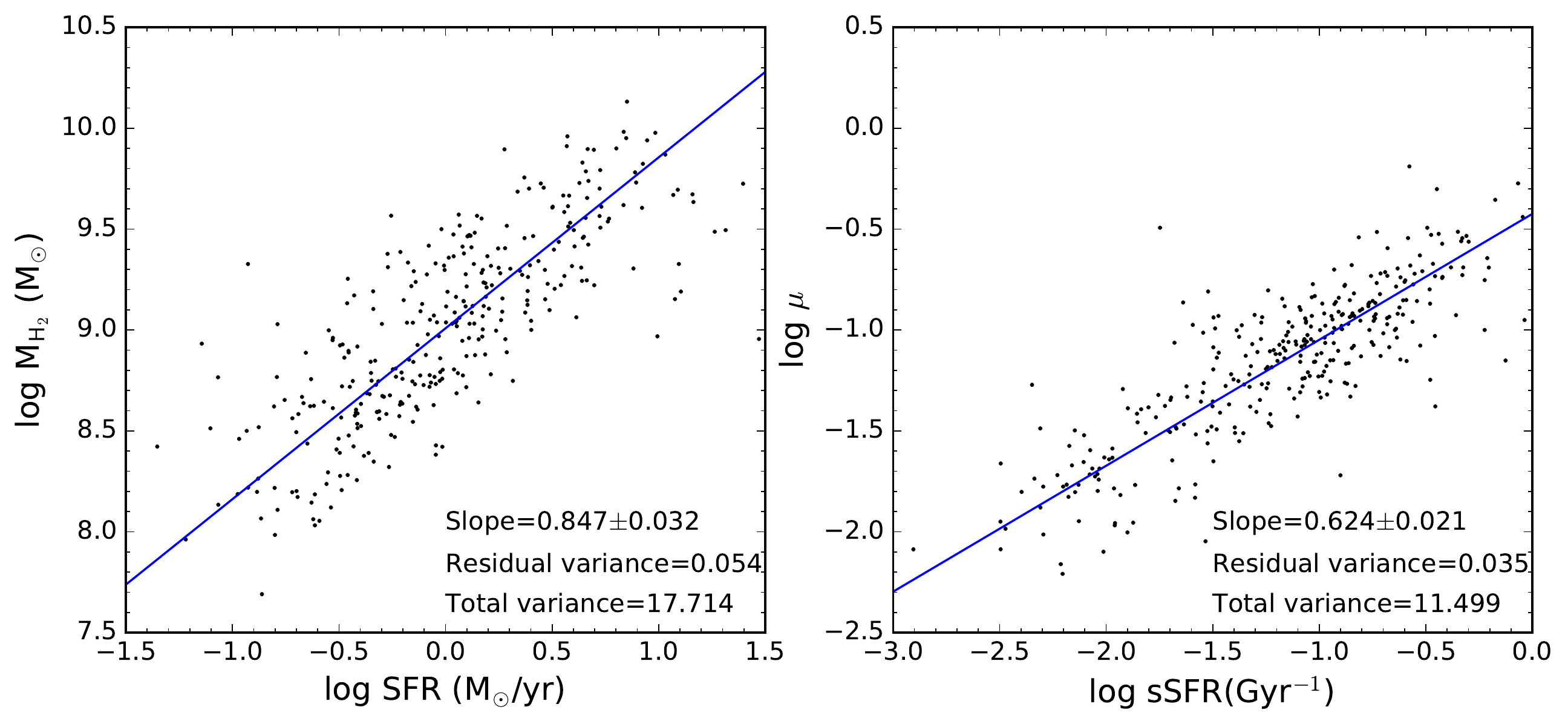}
    \end{center}
\caption{Comparison of the $M_{\rm H_2}$-SFR relation (left) and $\mu$-sSFR relation (right). Black dots are individual galaxies. The blue solid lines show the best fits to the data using ODR fitting method. The slope, residual variance and total variance of the best-fitting model are given in the legend.}
 \label{Two_relations}
\end{figure*}

The left panel of Figure \ref{Two_relations} shows the molecular gas mass $M_{\rm H_2}$ as a function of SFR, which is often referred as the integrated KS law \citep{2014ApJ...793...19S,2016A&A...590A..27G,2018MNRAS.479..703C,2019arXiv191207633B}. The right panel shows the molecular gas to stellar mass ratio $\mu=M_{\rm H_2}/M_{*}$ as a function of sSFR. Given sSFR $=\mu \times $SFE, the $\mu$-sSFR relation is closely related to the SFE-sSFR relation or the $\tau$-sSFR relation. The $\mu$-sSFR, SFE-sSFR and $\tau$-sSFR relations have been shown in many studies from local to high redshift (see the references in the introduction).

Black dots in both panels are individual galaxies and blue solid lines are the best fits to the data using the orthogonal distance regression (ODR) fitting method, which calculates the sum of the orthogonal distances from the data points to the fitted line. Since ODR accounts for variances on both the x- and the y- axis, we use ODR instead of the traditional least square method which measures variances only parallel to the y-axis. The slope, residual variance and total variance of the best-fitting model are noted in the legend. The total variance is the sum of squares of the variances in both x- and y-directions. The residual variance is the total variance divided by the degrees-of-freedom in the data. The smaller the residual variance and the total variance, the tighter the correlation between the y-axis and x-axis variables. The logarithmic slope of the $M_{\rm H_2}$-SFR relation in the left panel is $\sim$ 0.847, in good agreement with the relation $M_{\rm H_2} \propto \rm SFR^{0.81}$ as in \citet{2014ApJ...793...19S}. We also note that the exact value of the slope strongly depends on the sample and the adopted $\alpha_{\rm CO}$. For instance, including more starbursts and ULIRGs and using a smaller $\alpha_{\rm CO}$ \citep{2004ApJ...606..271G,2015ApJ...805...31L}, or including more low-metallicity dwarf galaxies at the low SFR end and using a much larger $\alpha_{\rm CO}$ \citep{2009ApJ...696.1834W,2019ApJ...872...16D}, will both produce a much shallower slope of the $M_{\rm H_2}$-SFR relation. Since our primary goal is to compare the relative differences between different scaling relations, their absolute slopes and normalizations are less critical to our analysis.

We also estimate the 1$\sigma$ scatter, which is 0.23 dex for the $M_{\rm H_2}$-SFR relation and 0.19 dex for the $\mu$-sSFR relation. The typical measurement errors as estimated by xCOLD GASS \citep{Saintonge:2017iz}, is 0.18 dex in $M_{\rm H_2}$ and 0.13 dex in SFR. The typical 1$\sigma$ measurement errors in $M_{*}$ derived from SDSS is about 0.1 dex. Hence the errors in $\mu$ and sSFR is about 0.21 dex and 0.16 dex, respectively, estimated from error propagation. Interestingly, even if the measurement errors of $\mu$ and sSFR are both larger than those of $M_{\rm H_2}$ and SFR, the $\mu$-sSFR relation has a smaller scatter than the $M_{\rm H_2}$-SFR relation. In fact, the combined measurement errors of $\mu$ and sSFR on the orthogonal direction to the fitted line is 0.2 dex. This means the scatter of the $\mu$-sSFR relation, 0.19 dex, can be entirely explained by the measurement errors of $\mu$ and sSFR. This hence suggests that there is little room to further reduce the scatter of this relation by including other galaxy properties. In other words, the $\mu$-sSFR relation will most likely show no dependence on other galaxy properties, which is indeed the case as will be shown in later sections.

%-----------------------------------------------------------------------------------------------------------------------------
\subsection{Interrelationship between $M_{\rm H_2}$-SFR and $\mu$-sSFR}

It is evident that the residual variance and total variance of the $\mu$-sSFR relation are significantly smaller than that of the $M_{\rm H_2}$-SFR relation, i.e. than that of the integrated KS law. Since SFR = sSFR$\times M_{*}$ and $M_{\rm H_2}=\mu \times M_{*}$, the $M_{\rm H_2}$-SFR relation can be simply transformed into the $\mu$-sSFR relation via dividing by $M_{*}$. This means that if each galaxy in the left panel of Figure \ref{Two_relations} is shifted by its log $M_{*}$ in both the x-axis and y-axis, we obtain the right panel.

Will this shift change the overall variance (or scatter) of the relation? We first consider one special case. If all the galaxies would have the same stellar mass, then the amount of shift will be the same for each galaxy. In this case, the shift will change neither the overall variance of the relation nor its slope, i.e. the total variance and residual will be the same for $M_{\rm H_2}$-SFR relation and $\mu$-sSFR relation. In reality, since galaxies have different stellar masses, the amount of shift will be different for different galaxies. Therefore, when we transform the $M_{\rm H_2}$-SFR relation into the $\mu$-sSFR relation by shifting each galaxy according to its stellar mass, its slope will change, and its scatter can increase or decrease.

So the question to ask is what is the probability if this transformation (by shifting each galaxy) would decrease the scatter? If the probability is high, then the fact that the $\mu$-sSFR relation has a smaller scatter as shown in Figure \ref{Two_relations} could be due to some random effects and does not imply any important physical reason.

\begin{figure}[htbp]
    \begin{center}
       \includegraphics[width=\columnwidth]{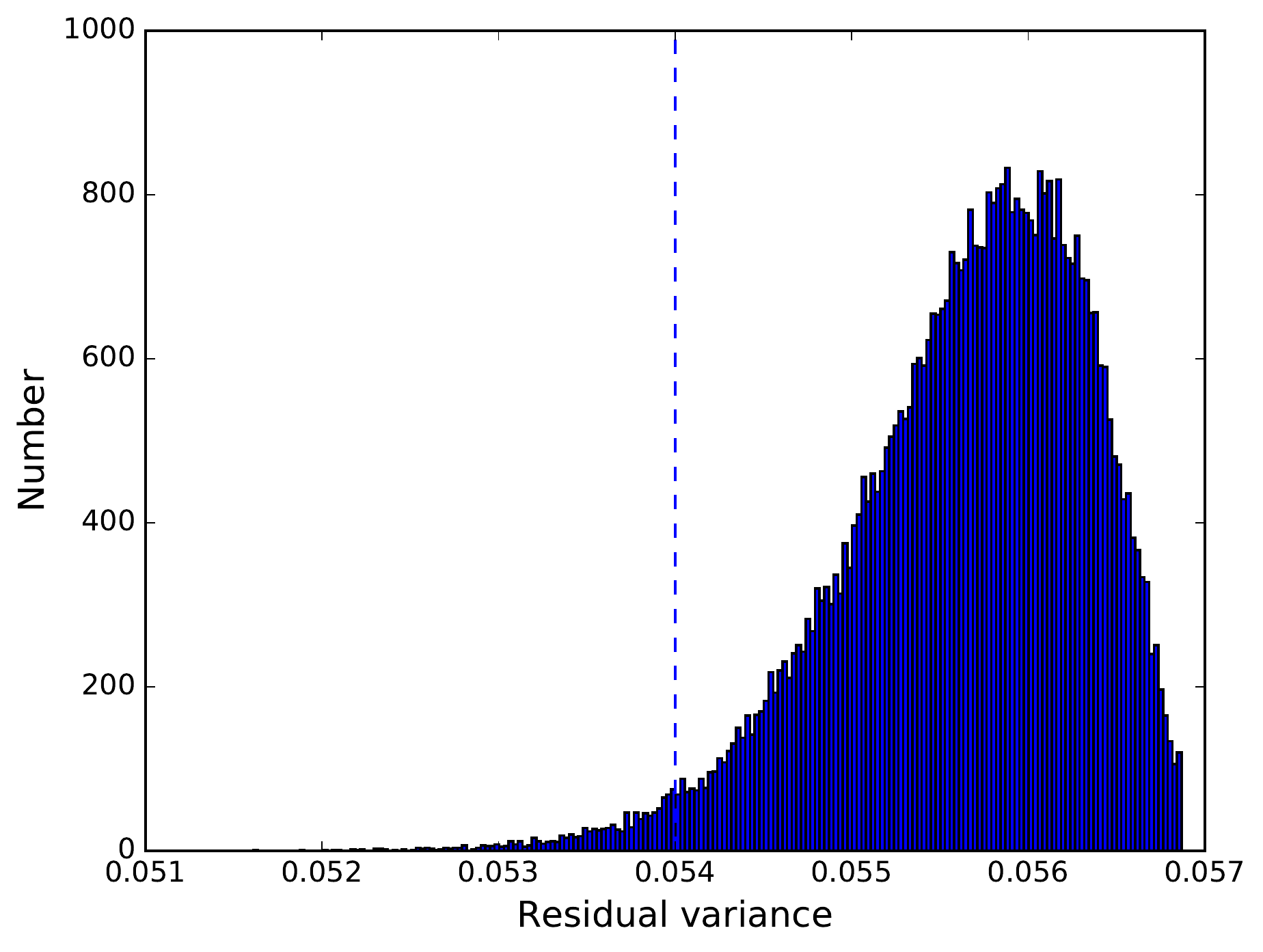}
    \end{center}
\caption {The probability distribution of the residual variance estimated from Monte Carlo Simulations with 50000 realizations. In each realization, we first shuffle the $M_{*}$ of all galaxies randomly, shift each galaxy in the left panel of Figure \ref{Two_relations} by its log $M_{*}$, and the $M_{\rm H_2}$-SFR plot is transformed into the $\mu$-sSFR plot. Then we do the ODR fitting and calculate the residual. The residual variance of the original integrated KS law (left panel in Figure \ref{Two_relations}) is marked by the vertical dashed line. The residual of the observed $\mu$-sSFR relation, 0.035, is significantly offset from the main distribution. The probability to achieve such a small residual variance, by shifting each galaxy with a random stellar mass, is extremely close to zero.}
 \label{shuffle}
\end{figure}

We use a simple Monte Carlo simulation to evaluate such a probability. In each realization, we first shuffle the stellar masses of all galaxies randomly, shift each galaxy in the left panel of Figure \ref{Two_relations} by its new log $M_{*}$, and then we do the ODR fitting and calculate the residual. We have done 50000 realizations and the distribution of the residual value is shown in Figure \ref{shuffle}.

There are two important facts. First, the residual value of the original $M_{\rm H_2}$-SFR relation, marked by the vertical dashed line, is significantly smaller than the mean or median of the distribution. This implies the tightness of the $M_{\rm H_2}$-SFR relation, i.e. the integrated KS law, is unlikely due to some random effects, but driven by some physical processes or reasons (see also the discussion about the more general experiments in the next paragraph). Second, the residual of the observed $\mu$-sSFR relation, 0.035, as noted in the right panel of Figure \ref{Two_relations}, is significantly offset from the main distribution. The probability to achieve such a small residual value, by shifting each galaxy with a random stellar mass, is extremely close to zero. This hence suggests \textit{the fact that the $\mu$-sSFR relation is tighter than the $M_{\rm H_2}$-SFR relation is driven by some more universal physical process or reason that controls the amount of shift of each galaxy in a remarkably precise way to achieve such a small residual value, and the amount of shift happens to be the stellar mass of each galaxy}. This also implies that stellar mass is not a random variable, but correlates with $M_{\rm H_2}$ and with SFR as well, i.e. stellar mass plays an important role in linking star formation with the molecular gas content. We will further discuss the role of stellar mass in the next section.

In more general experiments, we shift each galaxy according to not only its stellar mass (as above), but also its other properties such as size, B/T, metallicity, etc., to get different scaling relations. Among all these relations (including the original $M_{\rm H_2}$-SFR relation, i.e. do not shift), the $\mu$-sSFR has the smallest scatter.  To put it another way, on the $M_{\rm H_2}$-SFR plane, if we shift each galaxy by an amount that is equal to one of its properties to get the smallest scatter, that amount is precisely log $M_{*}$ and the result is the $\mu$-sSFR relation.

We note that the global slope (i.e. the slope of all galaxies) of the $\mu$-sSFR relation is shallower than the $M_{\rm H_2}$-SFR relation. In general, if the total ODR variance of two relations are the same, the one with a shallower slope is less correlated. An extreme example is a relation with zero slope, which means there is no correlation. We hence calculate the Pearson correlation coefficient \textit{p} (\textit{p} = 1 means total positive linear correlation, and \textit{p} = 0 means no linear correlation) and find that \textit{p} = 0.79 for $M_{\rm H_2}$-SFR relation and \textit{p} = 0.83 for $\mu$-sSFR relation. Therefore, even if the $\mu$-sSFR relation has a shallower slope, the correlation is still stronger than the $M_{\rm H_2}$-SFR relation, which requires the $\mu$-sSFR relation to have an even smaller variance (because of its shallower slope).  We will show in the next section that, in fact, the intrinsic slope of the $M_{\rm H_2}$-SFR relation (i.e. slope at a given stellar mass) is the same as the $\mu$-sSFR relation.

%-----------------------------------------------------------------------------------------------------------------------------
\subsection{Internal structure of $M_{\rm H_2}$-SFR and $\mu$-sSFR relations}

%---------------------------------------------------------------------------------------------------------------------
\begin{figure*}[htbp]
    \begin{center}
       \includegraphics[width=180mm]{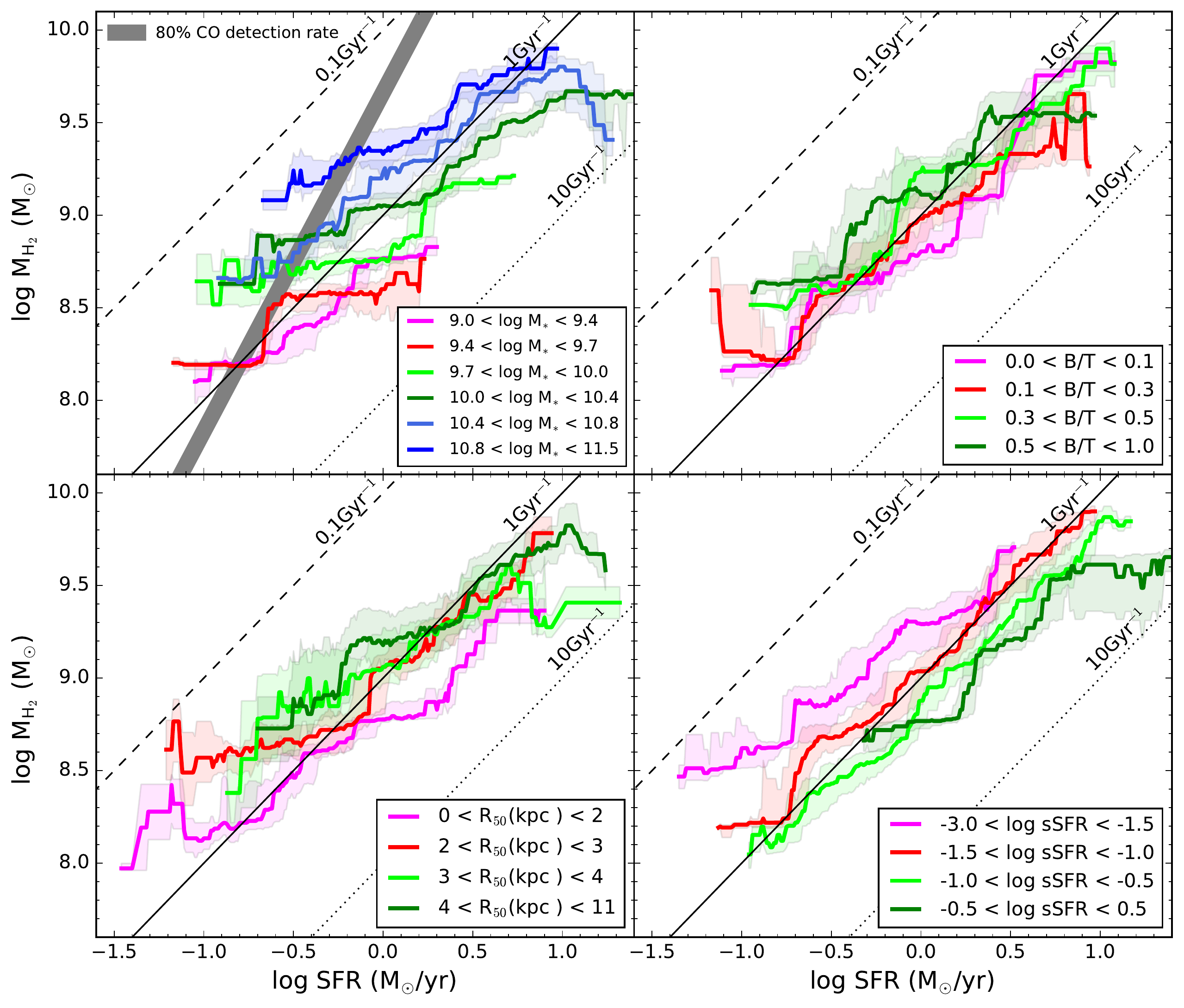}
    \end{center}
\caption{The average molecular gas mass as a function of SFR in different stellar mass, bulge-to-total ratio, effective radius and sSFR. The average values are calculated with a sliding box of 0.5 dex in SFR. Error bars on each line indicate the 32th and 68th percentile of the galaxy distribution. The black diagonal lines are the constant star formation efficiency in three different values. The thick gray line marks the CO detection ratio of $\sim$80\%. On the right hand side of the gray line, the average detection ratio is larger than 80\% and the results in this part should be reliable against selection effects (see text and Appendix A for detail).}
 \label{MH2-SFR}
\end{figure*}

%---------------------------------------------------------------------------------------------------------------------
\begin{figure*}[htbp]
    \begin{center}
       \includegraphics[width=180mm]{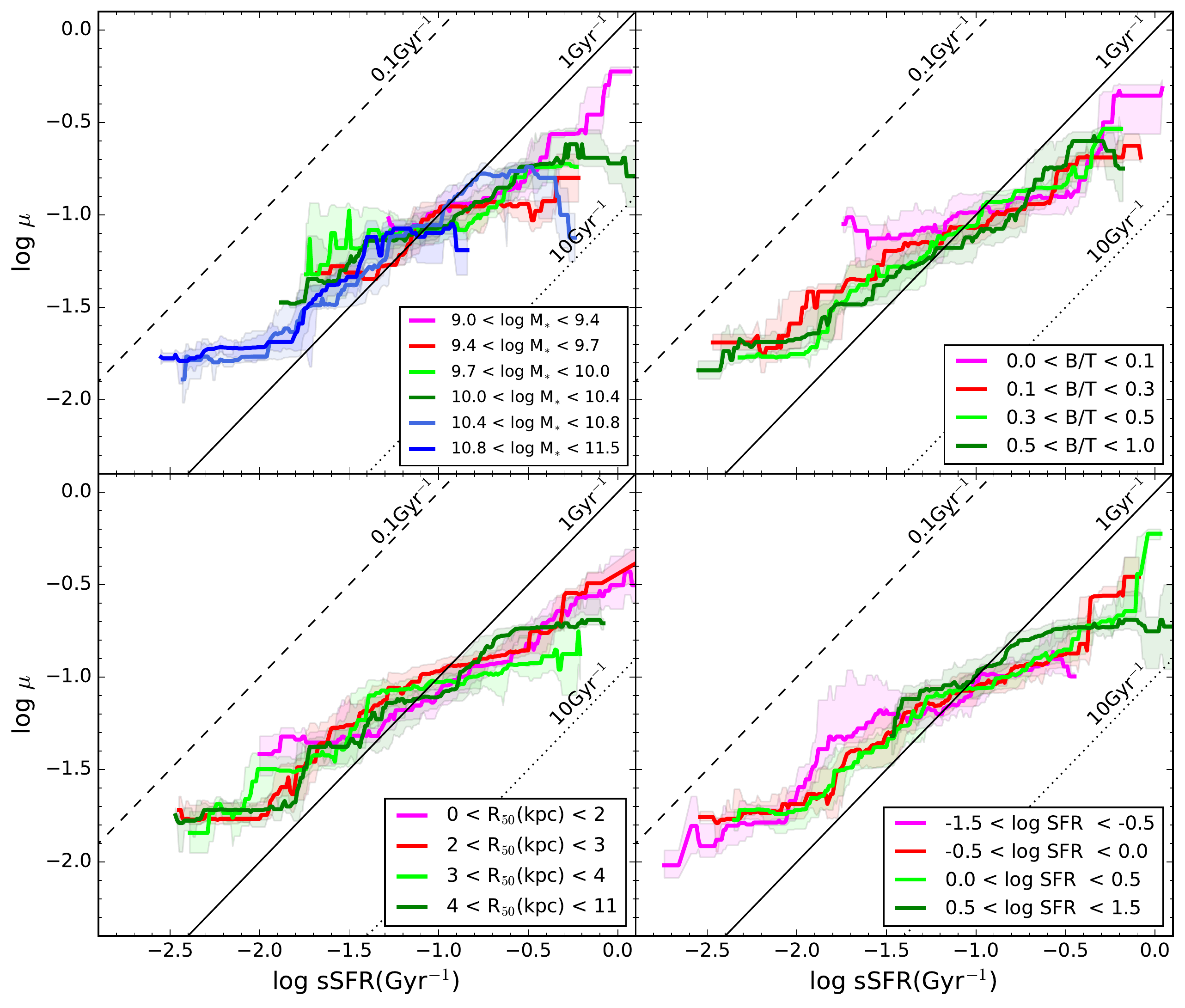}
    \end{center}
\caption{As for Figure \ref{MH2-SFR}, but for the $\mu$-sSFR relation and the lower right panel is for different SFR bins. As discussed in the text and Appendix A, for galaxies with log sSFR $>$ -1.6 Gyr$^{-1}$, the average CO detection ratio is larger than 80\%, where the results should be reliable.}
 \label{u-ssfr}
\end{figure*}

To further explore why the $\mu$-sSFR relation has a significantly smaller scatter than the $M_{\rm H_2}$-SFR relation and the underlying physics of star formation and quenching, we analyze their dependence on other key galaxy properties, by splitting the galaxies into different bins of stellar mass, bulge-to-total ratio, effective radius, sSFR (for $M_{\rm H_2}$-SFR) and SFR (for $\mu$-sSFR). The results are shown in Figure \ref{MH2-SFR} (for $M_{\rm H_2}$-SFR) and Figure \ref{u-ssfr} (for $\mu$-sSFR). The median values are calculated with a sliding window of 0.5 dex in SFR (Figure \ref{MH2-SFR}) and sSFR (Figure \ref{u-ssfr}), with a minimum of 4 galaxies inside each window. The black diagonal lines (with slope of unity) in both Figure \ref{MH2-SFR} and \ref{u-ssfr} indicate the constant SFE for three different values (given that SFR = SFE $\times M_{\rm H_2}$ and sSFR = SFE $\times \mu$).

As discussed in Appendix A, the CO detection ratio strongly depends on the SFR and sSFR. In the upper left panel of Figure \ref{MH2-SFR}, the thick gray line marks the detection ratio of $\sim$ 80\% (except for galaxies with log $M_{*} < $ 9.4). On the right hand side of the gray line, the detection ratio is larger than 80\% and the results in this part should be reliable against selection effects. As shown in Figure A1, below log $M_{*} \sim$  9.4, the average detection ratio is slightly lower than for more massive galaxies with similar sSFR.  The detection ratio as a function of sSFR is shown in the right panel of Figure A2. In Figure \ref{u-ssfr}, for galaxies with log sSFR $>$ -1.6 Gyr$^{-1}$, the detection ratio is larger than 80\%, where the results should be reliable.  We also discuss in Appendix A, even for galaxies below the 80\% detection limit, i.e. in the low SFR regions with rapidly increasing numbers of non-detections in CO, why our results might be still meaningful and not significantly biased by the strong selection effect in CO.  As discussed in Appendix A, if we exclude elliptical galaxies whose SFR could be overestimated, the CO detection ratio at the low SFR end will significantly increase, since about 50\% of the galaxies without CO detections are ellipticals, which are presumably quenched at high redshift with little remaining cold gas and ongoing SFR. They may also follow different scaling relations than those presented in this paper. Indeed, for galaxies with CO detections, the ellipticals are only about 5\%.  On the other hand, the selection effect will affect all scaling relations explored in this paper. Our analyses are primarily comparing the relative differences, for instance, between the $M_{\rm H_2}$-SFR and $\mu$-sSFR relations, not to determine their absolute slope and normalization, hence selection effects are less critical.

Comparing Figure \ref{MH2-SFR} and \ref{u-ssfr}, the $M_{\rm H_2}$-SFR relation shows systematic dependence on $M_{*}$, B/T, R$_{50}$ and sSFR, and in particular, strong systematic dependence on $M_{*}$. On the contrary, the $\mu$-sSFR relation shows almost no dependence on these parameters (note the lower right panel in Figure \ref{u-ssfr} is for different SFR bins). This explains why the $\mu$-sSFR relation has a smaller scatter (in terms of residual variance and total variance) than the $M_{\rm H_2}$-SFR relation.

\subsubsection{$M_{\rm H_2}$-SFR and $\mu$-sSFR on $M_{*}$}

The strong dependence of the $M_{\rm H_2}$-SFR relation on stellar mass as shown in Figure \ref{MH2-SFR} is also consistent with the so-called extended KS relation, which finds the scatter of the original KS relation can be further reduced by incorporating stellar mass into the fitting \citep{1985ApJ...295L...5D,1994ApJ...430..163D,2011ApJ...733...87S,2018ApJ...853..149S}.

\begin{figure*}[htbp]
    \begin{center}
       \includegraphics[width=180mm]{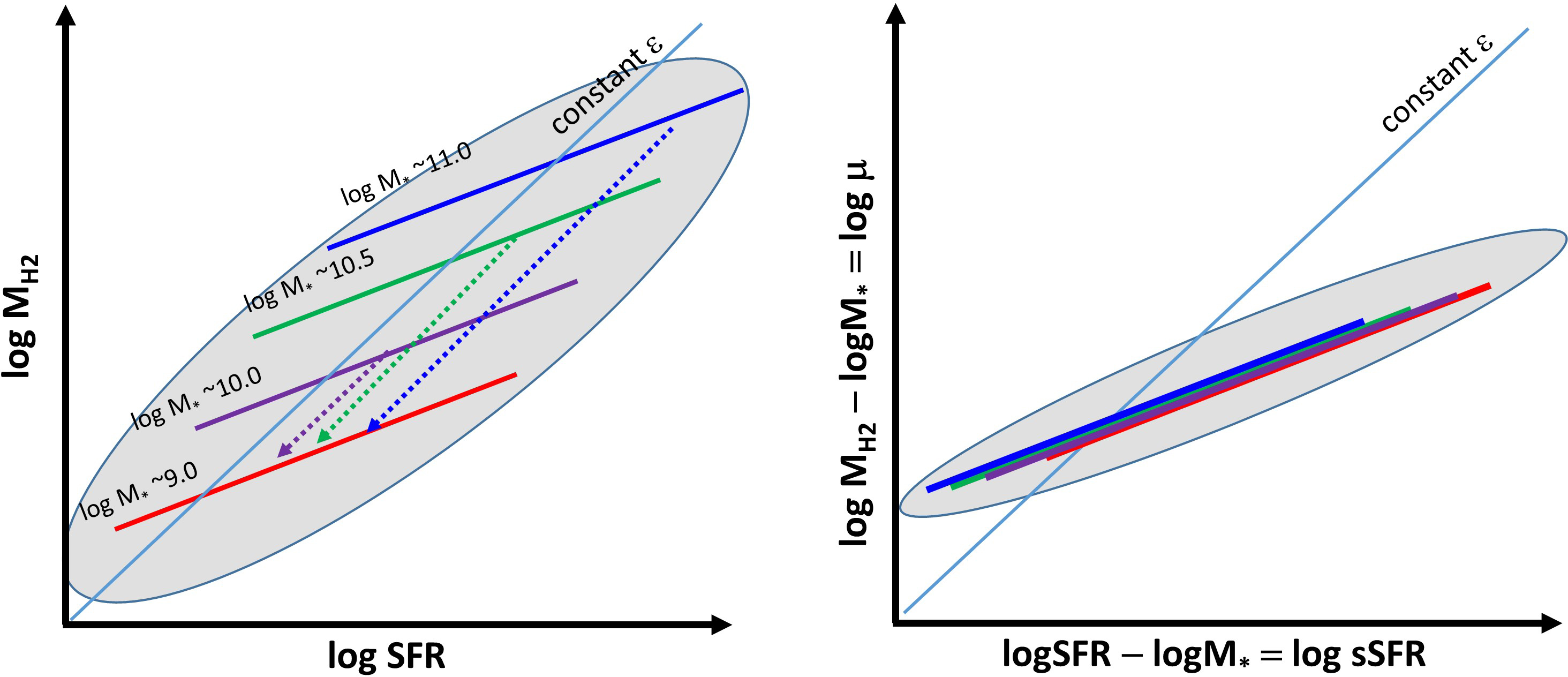}
    \end{center}
\caption {Illustration of the interrelationship between the $M_{\rm H_2}$-SFR (i.e. the integrated KS law) and $\mu$-sSFR relation, and their dependence on stellar mass. The $M_{\rm H_2}$-SFR relation can be transformed into the $\mu$-sSFR relation by shifting each galaxy by its log $M_{*}$. For a given stellar mass bin (i.e. galaxies with similar $M_{*}$), the shift is the same. Therefore, when the $M_{\rm H_2}$-SFR relation at a given stellar mass bin is transformed into the $\mu$-sSFR relation, it is shifted along the diagonal line (i.e. along the constant SFE line, as the shift is -log $M_{*}$ in both x-axis and y-axis) and its slope remains unchanged. In other words, the intrinsic slope of the $M_{\rm H_2}$-SFR relation (i.e. slope at a given stellar mass) is the same as the $\mu$-sSFR relation. The global $M_{\rm H_2}$-SFR relation (shaded region in the left panel) is a collection and superposition of the $M_{\rm H_2}$-SFR relations at different stellar masses that all have the same slope, but different zero points depending on the stellar mass. The $\mu$-sSFR relation (right panel) at different stellar masses that all have the same slope and the same zero point, hence a much smaller overall scatter. See the text for more discussion.}
 \label{transformation}
\end{figure*}

Here we further explore why incorporating stellar mass can significantly reduce the scatter. We first look at the upper left panel of Figure \ref{MH2-SFR}. Interesingly, the $M_{\rm H_2}$-SFR relation at different stellar mass bins are all parallel to each other with a similar slope. The slope of individual lines is in fact the same as the slope of the $\mu$-sSFR relation (of both the global one and stellar mass binned ones as shown in the upper left panel of Figure \ref{u-ssfr}), which is $\sim$ 0.6 as shown in the right panel of Figure \ref{Two_relations}. As discussed before, the $M_{\rm H_2}$-SFR relation can be transformed into the $\mu$-sSFR relation by shifting each galaxy by its log $M_{*}$. For a given stellar mass bin (i.e. galaxies with similar $M_{*}$), the shift is the same. Therefore, when we transform the $M_{\rm H_2}$-SFR relation at a given stellar mass bin into the $\mu$-sSFR relation, we shift it along the diagonal line (i.e. along the constant SFE line, as the shift is -log $M_{*}$ in both x-axis and y-axis) and its slope remains unchanged. This process is illustrated in Figure \ref{transformation}.

We stress that, as above, the slope of the $M_{\rm H_2}$-SFR relation at different stellar mass bins is the same (also the same as the slope of the $\mu$-sSFR relation, about 0.6), which is much shallower than the slope of the global $M_{\rm H_2}$-SFR relation ($\sim$ 0.8). Therefore, the global $M_{\rm H_2}$-SFR relation is a collection and superposition of the $M_{\rm H_2}$-SFR relations at different stellar masses (that all have the same slope, but different zero points depending on the stellar mass). The $\mu$-sSFR relations at different stellar masses all have the same slope and the same zero point, hence the global $\mu$-sSFR relation has a much smaller overall scatter and is a more universal relation than the $M_{\rm H_2}$-SFR relation.

Putting it another way, how can we explain simultaneously why (a) the slopes of the $M_{\rm H_2}$-SFR at a given (different) M$_{*}$ are all the same and (b) the separation between the different $M_{\rm H_2}$-SFR relations in different M$_{*}$ bins is controlled by M$_{*}$? Both (a) and (b) must be satisfied, otherwise when the $M_{\rm H_2}$-SFR relations at different M$_{*}$ bins are transformed into the $\mu$-sSFR relation, the different lines will not collapse into one single relation, as observed. A priori, it is not completely obvious why conditions (a) and (b) should hold, as observed. Also Figure \ref{shuffle} has ruled out reason due to some coincidence or random effect. One natural explanation is that the $M_{\rm H_2}$-SFR relation is actually originated from the $\mu$-sSFR relation.

\subsubsection{$M_{\rm H_2}$-SFR and $\mu$-sSFR on B/T, R$_{50}$ and sSFR(SFR)}

As mentioned before, the $M_{\rm H_2}$-SFR relation also shows systematic dependence on B/T, R$_{50}$ and sSFR, though not as significant as its dependence on $M_{*}$.

The upper right panel in Figure \ref{MH2-SFR} shows the galaxies with a larger B/T, on average, have a lower SFE, though the difference of SFE at different B/T is quite small. This trend is consistent with the results shown in \citet{2012ApJ...758...73S,2016MNRAS.462.1749S} that the bulge-dominated galaxies tend to convert gas into stars less efficiently. This can be explained by morphological quenching \citep{2009ApJ...707..250M,2014ApJ...785...75G,2020MNRAS.495..199G} that the presence of a central stellar bulge can stabilize the gas on the disk and reduce the SFE. The lower left panel shows that the galaxies with a smaller R$_{50}$, on average, have a higher SFE. This can be explained as at a given $M_{\rm H_2}$, the galaxies with a smaller R$_{50}$ have a higher average gas density (by assuming the \htwo\ gas distribution follows the stars), hence may have a higher SFE.

The $\mu$-sSFR relation, on the contrary, shows little or no dependence on $M_{*}$, B/T, R$_{50}$, SFR and other galaxy properties we have explored, including gas-phase metallicity and environment (in terms of overdensity and central/satellite dichotomy). As mentioned before, given sSFR = $\mu \times$SFE, the $\mu$-sSFR relation is also closely related to the SFE-sSFR (or equivalently the $\tau$-sSFR) relation, which will be explored in the next section.

The fact that $M_{\rm H_2}$-SFR relation shows systematic dependence on $M_{*}$, B/T, R$_{50}$ and sSFR implies that the SFE systematically depends on $M_{*}$, B/T, R$_{50}$ and sSFR (Figure \ref{MH2-SFR}). On the other hand, at a given sSFR, the SFE does not depend on $M_{*}$, B/T, R$_{50}$ and SFR (Figure \ref{u-ssfr}). Putting together, in the local Universe, the SFE (or equivalently the gas depletion timescale $\tau$) is primarily correlated with the sSFR, not with SFR or with $M_{\rm H_2}$.

It should be noted that this does not suggest the stellar bulge and/or the gas density would have no effect on the SFE. This is because although SFE is primarily correlated with the sSFR, sSFR may depend on B/T and/or the gas density (and/or on other galaxy properties, e.g., on $M_{*}$). Therefore, one of the next key question to explore is what determines the sSFR and drives its evolution. \citet{Peng:2014hn}, the first paper in this series, explored the importance of the sSFR and stressed its critical role in driving galaxy evolution.

Indeed, the important fact is that, as shown in Figure \ref{Two_relations} (right panel) and Figure \ref{u-ssfr}, the unique feature of $\mu$-sSFR relation (its tightness, a single slope and independence of other key galaxy properties) remains unchanged across the entire observed range of sSFR, from star-bursting to galaxies in the process of being quenched. This suggest that \textit{the same star formation physics may operate across the entire galaxy population, for galaxies with enormous differences in stellar masses, size and structure, from star-bursting to galaxies in the process of being quenched. All galaxies just evolve on the same scaling relation of $\mu$-sSFR, as their sSFR decrease due to secular evolution with cosmic time or quenching.} We will further explore this aspect in our next paper.

%-----------------------------------------------------------------------------------------------------------------------------
\section{Interrelationships between $M_{\rm H_2}$-$M_{*}$, SFR-$M_{*}$ and SFE-sSFR, SFE-$\mu$}

%---------------------------------------------------------------------------------------------------------------------
\begin{figure*}[htbp]
    \begin{center}
       \includegraphics[width=180mm]{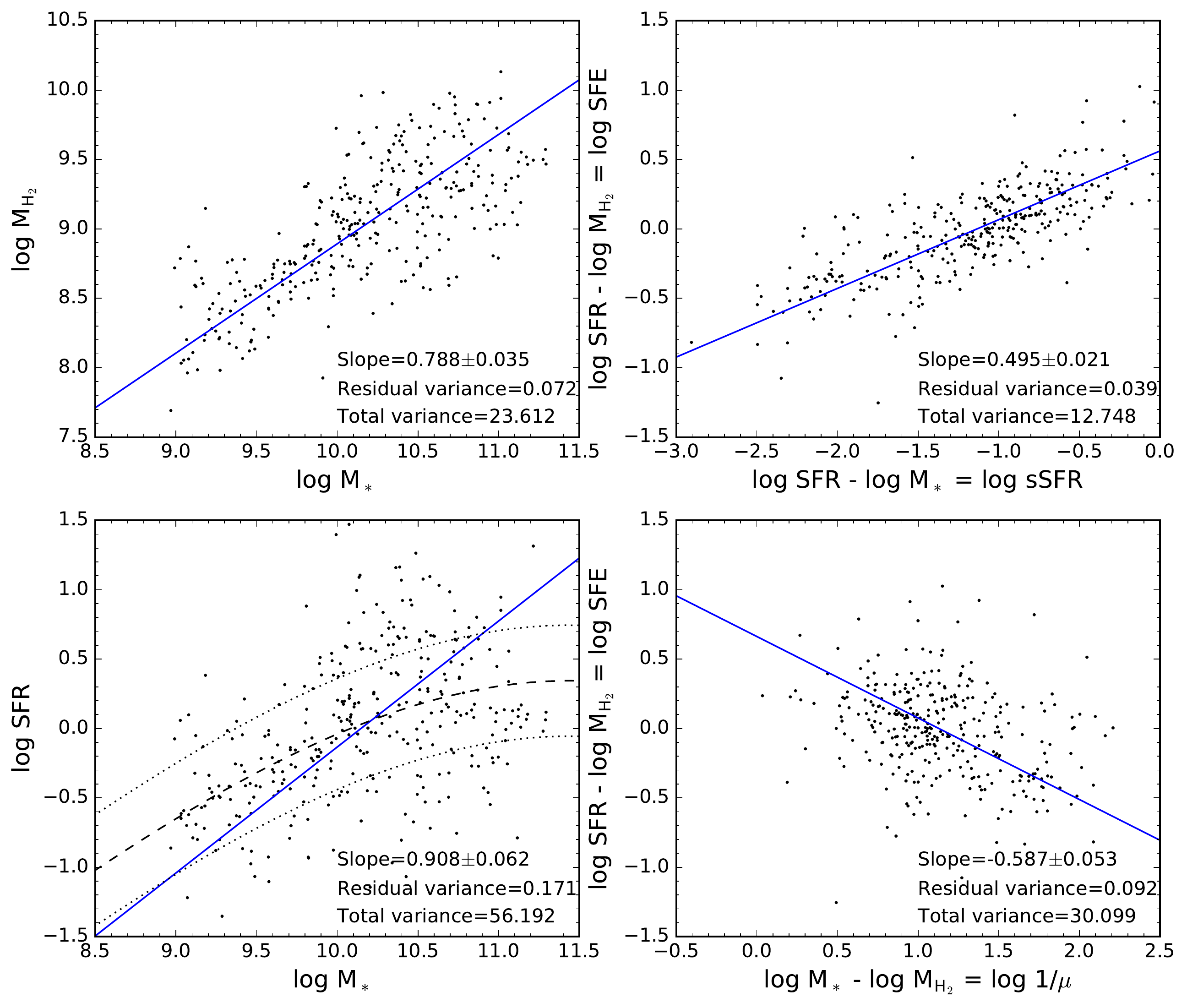}
    \end{center}
\caption{Comparison of the $M_{\rm H_2}$-$M_{*}$ relation (upper left) and SFE-sSFR relation (upper right); SFR-$M_{*}$ relation (bottom left) and SFE-1/$\mu$ relation (bottom right). Black dots are individual galaxies. The blue solid lines show the best fits to the data using ODR fitting method. The slope, residual variance and total variance of the best-fitting model are given in the legend. The dashed line indicates the position of the star-forming main sequence defined in \citet{2016MNRAS.462.1749S}. The dotted lines indicate $\pm$0.4 dex scatter around the main sequence.}
 \label{various_relations}
\end{figure*}

In addition to the star formation law, the other two important scaling relations widely discussed in the literature are the $M_{\rm H_2}$-$M_{*}$ and SFR-$M_{*}$ relations. As mentioned in the introduction (with the references there), for star-forming galaxies, these two relations are called as molecular gas main sequence (MG-MS) and star-forming main sequence (SF-MS). It should be noted that unlike many previous studies, in our analysis we do not differentiate between star-forming and passive galaxies. We include all galaxies from star-bursting ones and  down to those with the lowest observable SFR and \htwo \ gas mass. As we have shown in the previous section, we will show more evidence that the same star formation physics may operate across the entire galaxy population, from star-bursting to galaxies in the process of being quenched, as also argued in \citet{2018ApJ...853..179T,2020arXiv200306245T}.

Similar to the mutual transformation between $M_{\rm H_2}$-SFR and $\mu$-sSFR relations, the upper two panels in Figure \ref{various_relations} show that the $M_{\rm H_2}$-$M_{*}$ relation can be transformed into the SFE-sSFR relation by shifting each galaxy by its log SFR (and vice versa); the bottom two panels show that the SFR-$M_{*}$ relation can be transformed into the SFE-1/$\mu$ relation by shifting each galaxy by its log $M_{\rm H_2}$ (and vice versa). As dicussed in Section 3.1, in principle this shift can increase or decrease the overall scatter of the relation, and also change its slope. As indicated by the residual variance and total variance in the legend in each panel, the SFE-sSFR and SFE-1/$\mu$ relations have actually significantly smaller scatter than their corresponding $M_{\rm H_2}$-$M_{*}$ and SFR-$M_{*}$ relations.

As discussed before, given sSFR = $\mu \times$SFE, the $\mu$-sSFR relation is closely related to the SFE (or $\tau$)-sSFR relation. The combined measurement errors of SFE and sSFR on the orthogonal direction to the fitted line is 0.21 dex, which is similar to the scatter of the SFE-sSFR relation of 0.2 dex. Similar to the $\mu$-sSFR relation, this means the scatter of the SFE-sSFR relation can also be entirely explained by the measurement errors, and there is little room to further reduce the scatter by including other galaxy properties. Indeed, we find the SFE-sSFR relation also shows little or no dependence on other key galaxy properties, including $M_{*}$, B/T, R$_{50}$ and SFR, as shown in Figure A4 in Appendix B.

The $M_{\rm H_2}$-$M_{*}$ and SFR-$M_{*}$ relations evidently have the largest scatters than other discussed scaling relations. One apparent feature is that both $M_{\rm H_2}$-$M_{*}$ and SFR-$M_{*}$ relations consist of two structures (that are not lying on the same linear relation), one relatively tight sequence for star-forming galaxies (i.e. MG-MS for $M_{\rm H_2}$-$M_{*}$ relation and SF-MS for SFR-$M_{*}$ relation) and one loose cloud for passive galaxies, as illustrated in Figure \ref{FSFC}. The passive cloud is not entirely clear in the left two panels of Figure \ref{various_relations}, due to the relatively small size and sample selection of the xCOLD GASS. However, the existence of the passive cloud is naturally expected.

As discussed in \citet{2015ApJ...801L..29R}, the existence of a star-forming sequence and a passive cloud is primarily caused by quenching. Most galaxies in the local Universe are quenched through strangulation or starvation by halting the cold gas supply \citep{Peng:2015bq,Trussler_2019,2019ApJ...884L..52Z}. Therefore, when quenching starts, both $M_{\rm H_2}$ and SFR decrease at a speed which is controlled by the gas depletion timescale. On the $M_{\rm H_2}$-$M_{*}$ and SFR-$M_{*}$ plane, the galaxy in the process of quenching evolves mostly downwards, with some modest stellar mass increase during quenching. This hence brings the galaxy off the MG-MS and SF-MS, evolving towards the passive cloud, and hence significantly increase the scatter of the $M_{\rm H_2}$-$M_{*}$ and SFR-$M_{*}$ relations.

The SFE-1/$\mu$ relation, which can be transformed from the SFR-$M_{*}$ relation and vice versa, has a much smaller scatter than the SFR-$M_{*}$ relation. This is because, as discussed above and  illustrated in Figure \ref{FSFC}, the SFR-$M_{*}$ relation consists of two structures that are not lying on the same linear relation, while the SFE-1/$\mu$ relation appears to be a single sequence. Given that sSFR = $\mu \times$SFE, the SFE-1/$\mu$ relation hence determines the star formation level (i.e. sSFR) of the galaxy. A high sSFR could be due to a high $\mu$, or a high SFE, or both. By their definition, $\mu$ tells the relative amount of gas in the galaxy and SFE tells how efficient the galaxy can transform the gas into stars. In principle, $\mu$ and SFE could share no causal link and show no correlation. Therefore, the existence of a clear negative correlation between SFE and 1/$\mu$ (i.e. positive correlation between SFE and $\mu$) as in Figure \ref{various_relations} already gives insight into the star formation and quenching processes. On average, galaxies with a high SFE also have a larger $\mu$ (and hence a higher sSFR), and vice versa. The scatter around the SFE-1/$\mu$ relation suggests that different galaxies can have different star formation or quenching states. For instance, the decrease of sSFR can be caused by suppressing SFE (e.g., due to morphological quenching) or by the decrease of $\mu$ (e.g., due to strangulation, or by outflow driven by feedback, or by stripping due to environment effect), or by both. On the SFE-1/$\mu$ plane, quenching due to suppressing SFE will make the galaxy evolve downwards in the bottom right panel of Figure \ref{various_relations}, while quenching due to decrease of $\mu$ will make the galaxy evolve horizontally to the right. As above, the positive correlation between SFE and $\mu$ suggests that, on average, for the galaxy population, both quenching mechanisms are operating. We also note that the scatter of the SFE-1/$\mu$ relation is 0.3 dex, which is larger than the combined measurement errors of SFE and $\mu$ of 0.22 dex. This implies that galaxies are not quenched in the same way. As discussed above, the detailed quenching process (e.g., quenching primarily due to suppressing SFE or primarily due to decreasing $\mu$) can introduce additional scatters of the SFE-1/$\mu$ relation.

Finally, the main reason why $M_{\rm H_2}$-$M_{*}$ and SFR-$M_{*}$ relations have significantly larger scatters than the corresponding SFE-sSFR and SFE-1/$\mu$ relations is that the former two relations consist of two structures (i.e. a star-forming sequence and a passive cloud) that are not lying on the same linear relation. What about if we select only star-forming galaxies, i.e. what is the scatter of the MG-MS and SF-MS? If we use the same definition as \citet{2016MNRAS.462.1749S}, i.e. use the lower dotted line in Figure A1 to select star-forming galaxies, we find the scatters of resulting MG-MS and SF-MS are much smaller than the $M_{\rm H_2}$-$M_{*}$ and SFR-$M_{*}$ relations for the full sample, as expected, but are still larger than the corresponding SFE-sSFR and SFE-1/$\mu$ relations.  This is partially caused by galaxies with elevated SFRs and starbursting galaxies. These galaxies are located above the MG-MS and SF-MS (hence contribute to increase the scatter), but are still on the SFE-sSFR and SFE-1/$\mu$ relations.

%-----------------------------------------------------------------------------------------------------------------------------
\section{The Fundamental Formation Relation (FFR) and summary}

%-----------------------------------------------------------------------------------------------------------------------------
\subsection{Discussions}
%---------------------------------------------------------------------------------------------------------------------
\begin{figure*}[htbp]
    \begin{center}
       \includegraphics[width=180mm]{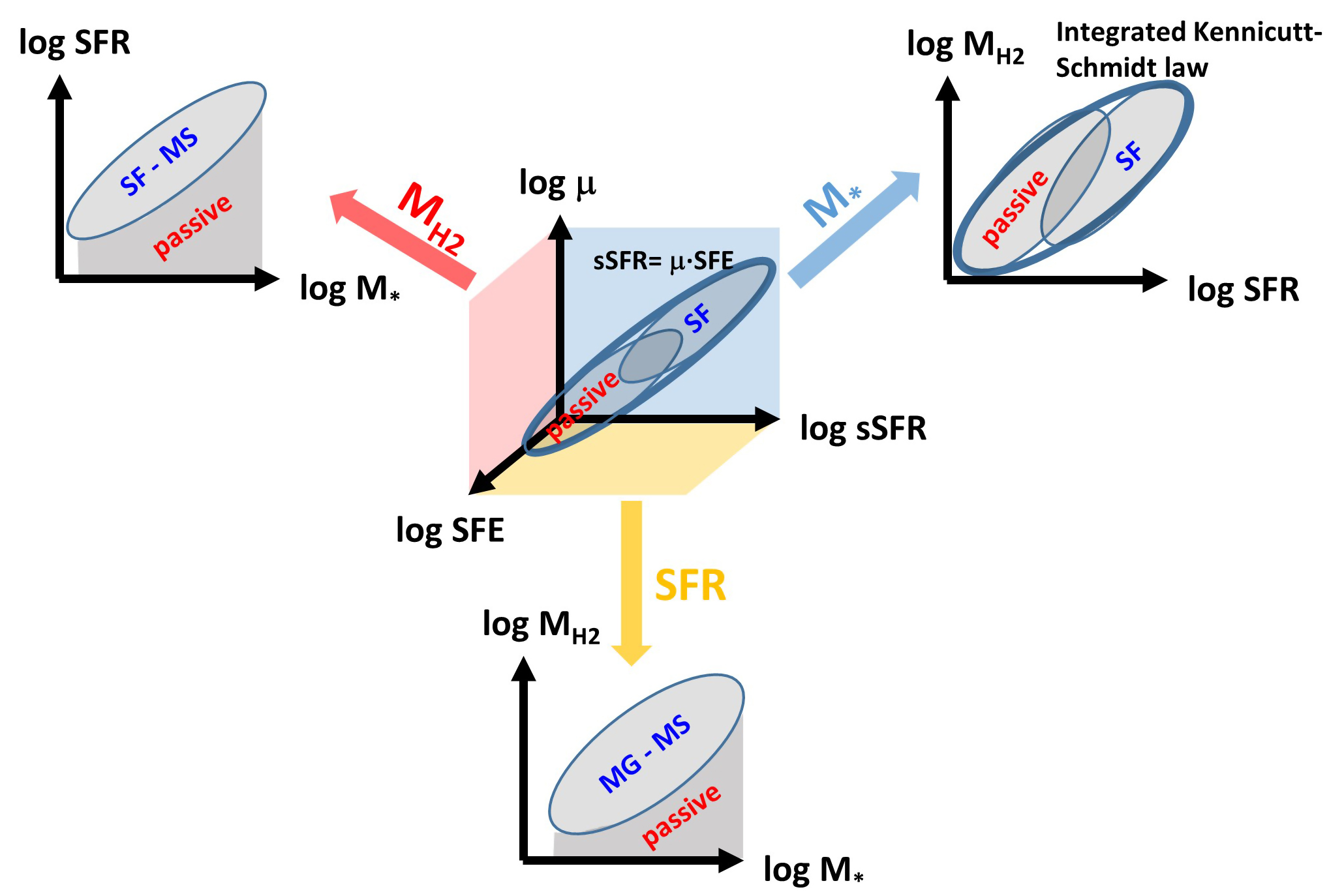}
    \end{center}
\caption{Illustration of the Fundamental Formation Relation (FFR), i.e. sSFR-$\mu$-SFE relation, and its relationship to the $M_{\rm H_2}$-SFR, $M_{\rm H_2}$-$M_{*}$ and SFR-$M_{*}$ relation. The $M_{\rm H_2}$-SFR relation is the integrated Kennicutt-Schmidt (KS) relation. For star-forming galaxies, $M_{\rm H_2}$-$M_{*}$ relation is called as the molecular gas main sequence (MG-MS) and SFR-$M_{*}$ relation is called as star-forming main sequence (SF-MS). The FFR provides a clean and simple framework to describe the evolution of galaxy, including both star formation and quenching.}
 \label{FSFC}
\end{figure*}

We have analyzed the two sets of key scaling relations in star formation and quenching: (1) absolute relations $M_{\rm H_2}$-SFR, $M_{\rm H_2}$-$M_{*}$ and SFR-$M_{*}$ and (2) specific relations $\mu$-sSFR, SFE (or $\tau$)-sSFR and SFE (or $\tau$)-$\mu$. In set (1), the $M_{\rm H_2}$-SFR is the integrated Kennicutt-Schmidt (KS) relation. For star-forming galaxies, $M_{\rm H_2}$-$M_{*}$ is called as the molecular gas main sequence (MG-MS) and SFR-$M_{*}$ is called as star-forming main sequence (SF-MS). We show the two sets of relations can be mutually tranformed, simply by shifting each galaxy by its log $M_{*}$, log SFR and log $M_{\rm H_2}$, respectively, as illustrated in Figure \ref{FSFC}. We find that the overall scatter of each relation in set (2) is significantly smaller than the corresponding one in set (1). We stress that since the amount of shift is different for different galaxies, the shift could increase or decrease the overall scatter of the relation. We show from a simple Monte Carlo simulation that the tightness of the $M_{\rm H_2}$-SFR relation, i.e. the integrated KS law, is unlikely due to some random effects. More interestingly, the probability to achieve an even smaller scatter as the observed $\mu$-sSFR relation is extremely close to zero, were it due to a random process. This is also true for the $M_{\rm H_2}$-$M_{*}$ v.s. SFE-sSFR relations, and SFR-$M_{*}$ v.s. SFE-$\mu$ relations. This suggests that the specific relations are much tighter than the absolute ones because of some more universal or important physical connection between these quantities. Also as discussed at the end of Section 3.2.1, the detailed connection between the $M_{\rm H_2}$-SFR and $\mu$-sSFR relation is beyond the fact that the latter relation has a smaller scatter, and further implies that the $M_{\rm H_2}$-SFR relation is actually originated from the $\mu$-sSFR relation.

We then show the significant smaller scatter of the specific relations is due to the systematic dependence of the relations in set (1) on other galaxies properties. For instance, $M_{\rm H_2}$- SFR relation shows systematic dependence on $M_{*}$, B/T, R$_{50}$ and sSFR, while the $\mu$-sSFR and SFE-sSFR relations show little or no dependence on $M_{*}$, B/T, R$_{50}$ and SFR. This implies that in the local Universe, the SFE (or $\tau$) is primarily correlated with the sSFR, not with SFR or with $M_{\rm H_2}$ or with $\delta_{\rm MS}$ (i.e. the SFR relative to that on the MS, defined as $\delta_{\rm MS}$ = SFR/SFR(MS). See the discussion on $\delta_{\rm MS}$ below).
%As discussed in Section 3.2.2, to understand the physics of the SFE-sSFR relation, we need to first understand what determines the sSFR and drives its evolution, which is closely linked to the dark matter halo accretion history \citep{Peng:2014hn,2014MNRAS.444.2071D}.

We stress that the unique feature of the $\mu$-sSFR and SFE-sSFR relations (i.e. their tightness, a single sequence with a single slope and independence of other key galaxy properties) remains unchanged across the entire observed range of sSFR, from star-bursting to galaxies in the process of being quenched. This suggests that the same star formation physics may operate across the entire galaxy population, for galaxies with enormous differences in stellar masses, sizes and structures. Therefore, the existence of two structures, i.e. a star-forming sequence (including the molecular gas main sequence and star-forming main sequence) and a passive cloud in the $M_{\rm H_2}$-$M_{*}$ and SFR-$M_{*}$ relations as illustrated in Figure \ref{FSFC}, does not necessarily mean that there are two different physical processes or reasons controlling star formation and quenching in a separate way.

What is the physics of the sSFR-$\mu$-SFE relation? First, the SFE (or $\tau$) is one of the key galaxy parameters and it describes how efficient the galaxy can form stars out of a given amount of cold gas. SFE (or $\tau$) is often related to galaxy's dynamical time that some fraction of molecular gas turned into stars per galactic orbital time under the gravitational instability of the cold gas on the disk, and the fraction depends on feedback physics \citep{1997ApJ...481..703S,1997ApJ...480..235E,1998ApJ...498..541K,2010MNRAS.407.2091G}. As further discussed in \citet{Genzel:2015fq}, the disk's orbital time is tied to the properties of the dark matter halo in the hierarchical clustering models for structure formation \citep{1998MNRAS.295..319M}. \citet{2020arXiv200306245T} further argues that $\tau$ is directly tied to the overall galactic clock, which is tied to cosmic time. Local processes may also be important or critical \citep[][for a review]{2020ARA&A..5812120S}, in particular in the local Universe where the average Toomre parameter of the disk is larger than unity \citep{Leroy:2008jk}, including local gas density, pressure, instability and turbulence that introduced by feedback or other dynamical process such as morphological quenching \citep{2009ApJ...707..250M,2014ApJ...785...75G}.

sSFR is another key parameter that describes how fast the galaxy can grow its stellar mass, since 1/sSFR is the e-folding timescale of the stellar mass. As discussed in \citet{Peng:2014hn} \citep[see also in][]{Lilly:2013ko,2014MNRAS.444.2071D}, the evolution of sSFR is primarily driven by the dark matter halo accretion history. Therefore, the observed tight correlation between SFE and sSFR, and its independence of most other galaxy properties, could be driven by the fact that both of them are closely related to the properties of the dark matter halo. It is also important to appreciate that 1/SFE and 1/sSFR are both timescales, one for gas and one for stars. In fact, \textit{$\tau$ (i.e. 1/SFE) is regarded as the primary parameter of gas evolution and is hence rephrased as "galactic clock" in \citet{2020arXiv200306245T}. While sSFR is regarded as the primary parameter of stellar population evolution in galaxies, which controls the mass function evolution of different galaxy populations, and is hence called as "cosmic clock" in \citet{2010ApJ...721..193P}.}

In the meanwhile, $\mu$ is mainly a measurement of the gravitational instability parameter Q for the gas disk \citep[e.g.,][]{2002ApJ...569..157W}. Therefore, the equation sSFR = SFE $\times \mu$ suggests that the star formation level (in terms of sSFR) is determined by the combination of the disk galactic dynamic timescale and gas instability.

It should be noted that in the sSFR-$\mu$-SFE relation, sSFR cannot be replaced by $\delta_{\rm MS}$. $\delta_{\rm MS}$ is often used in parallel with, or to substitute for sSFR in many studies, due to their similarity. As $\delta_{\rm MS}$ is simply the normalized quantity of sSFR (relative to the MS). Nevertheless, they are different in detail and in their physical meanings. sSFR has a unit of [1/time], while $\delta_{\rm MS}$ is dimensionless. If we replace sSFR with $\delta_{\rm MS}$ in Figure \ref{u-ssfr}, the different lines will not stay on a tight single sequence, and will become like Figure \ref{MH2-SFR}. This is because that the MS galaxies with different $M_*$ have different sSFR, and hence have difference $\mu$ and SFE according to the sSFR-$\mu$ and sSFR-SFE relations. Therefore, at the same $\delta_{\rm MS}$, for instance at $\delta_{\rm MS}=1$ (i.e. MS galaxies), the galaxies with different $M_*$ have different $\mu$ and SFE.

\subsection{Fundamental Formation Relation}

Putting together the key results above, we note that (1) the specific relations are much tighter than the absolute ones. In particular, the scatter of the $\mu$-sSFR and SFE-sSFR relations can be entirely explained by the measurement errors, which implies the intrinsic scatter of the $\mu$-sSFR and SFE-sSFR relations is extremely small. This is supported by their independence on other galaxies properties. (2) Unlike the absolute relations that contains two structures (i.e. a star-forming sequence and a passive cloud), the specific relations are single sequences, holding from star-bursting galaxies to the galaxies in the process of being quenched. This indicates that star formation in galaxies with distinct star formation status is regulated by the same star formation physics. (3) sSFR, SFE and $\mu$ are not only normalized quantities of $M_*$, SFR, $M_{\rm H_2}$, but they are also primary parameters in galaxy formation and evolution with specific physical meanings.

We therefore propose the sSFR-$\mu$-SFE(or $\tau$) relation as the Fundamental Formation Relation (FFR) and these three quantities are also linked by sSFR = $\mu \times$SFE. Other scaling relations, including integrated KS law, SF-MS and MG-MS, are derived from this fundamental cube, with a larger scatter.

In the FFR framework, galaxies with different stellar masses, sizes, structures, metallicity and in different environments, all evolve on the same single scaling relations of $\mu$-sSFR and SFE-sSFR (remember sSFR = $\mu \times$SFE). When their sSFRs change (no matter driven by secular evolution of the star-forming sequence or by quenching), their $\mu$ and SFE change according to the $\mu$-sSFR and SFE-sSFR relations. The SFE-$\mu$ relation, as discussed at the end of Section 4, reveals the deep causal connection between SFE and $\mu$; and clearly demonstrates, for instance, why a galaxy is quenching: is it due to a suppressed SFE, or due to the decrease of $\mu$, or both.

These unique features hence make the FFR an ideal framework to study galaxy formation and evolution. In our future papers, we will show various applications of the FFR in both observations and modelings, with further insight into the physics of the FFR.

%---------------------------------------------------------------------------------------------------------------------
\acknowledgments
We gratefully acknowledge the anonymous referee for comments and criticisms that have improved the paper. We thank Barbara Catinella, Sara L. Ellison, Li-Hwai Lin and Jing Wang for useful discussions. This work is supported in part by National Key R\&D Program of China (grant No. 2016YFA0400702, 2016YFA0400704, 2017YFA0402704 and 2017YFA0402703), the Natural Science Foundation of China (grants No.11773001, 11721303, 11991052, 11473002, 11690024, 11725313, 11861131007, 11420101002, 11733002 and 11633006), Chinese Academy of Sciences Key Research Program of Frontier Sciences (grant No.QYZDJSSW-SLH008 and QYZDJSSW-SYS008). A.R. acknowledges support from an INAF/PRIN-SKA 2017 (ESKAPE-HI) grant. F.M. acknowledges support from the INAF PRIN-SKA 2017 program 1.05.01.88.04. R.M. acknowledges ERC Advanced Grant 695671 "QUENCH'' and support from the Science and Technology Facilities Council (STFC). S.L. acknowledges support from the Swiss National Science Foundation.

%---------------------------------------------------------------------------------------------------------------------

\appendix

\section{Sample selection effects}

Since the xCOLD GASS sample is limited to the local Universe ($0.01<z<0.05$), the main selection effect is from the CO completeness limit, as mentioned in Section 2. The CO observation limit is set by the molecular gas to stellar mass ratio $M_{\rm H_2}/M_*$, which is 1.5\% for COLD GASS (i.e. $M_*>10^{10}\Msol$) and 2.5\% for COLD GASS-low (i.e. $10^{9}\Msol<M_*<10^{10}\Msol$). Only galaxies with reliable CO detections are used in our analysis in the main text, and non-detections with only upper limit in $M_{\rm H_2}$ have not been included.

To assess the significance of the selection effects on our results, we first show the \htwo\ detection ratio in xCOLD GASS sample on the SFR-$M_*$ plane in Figure A1. The dashed line indicates the position of the star-forming MS defined in \citet{2016MNRAS.462.1749S}. The dotted lines indicate $\pm$0.4 dex scatter around the main sequence, as in \citet{2016MNRAS.462.1749S} and \citet{Saintonge:2017iz}. The lower dashed line can be regarded as the approximate divide between star- forming galaxies and the galaxies that are quenched or in the process of being quenched. At a given stellar mass, the \htwo\ detection ratio starts to drop rapidly below this divide. The detection ratio also starts to drop modestly at log $M_{*} < $ 9.4, maybe due to the very low \htwo\ gas mass in the low stellar mass end. Also, the upper limits are less constraining at log $M_{*} < $ 9.4, as discussed in \citet{Saintonge:2017iz}.

\begin{figure}[htbp]
     \begin{center}
        \includegraphics[scale=0.5]{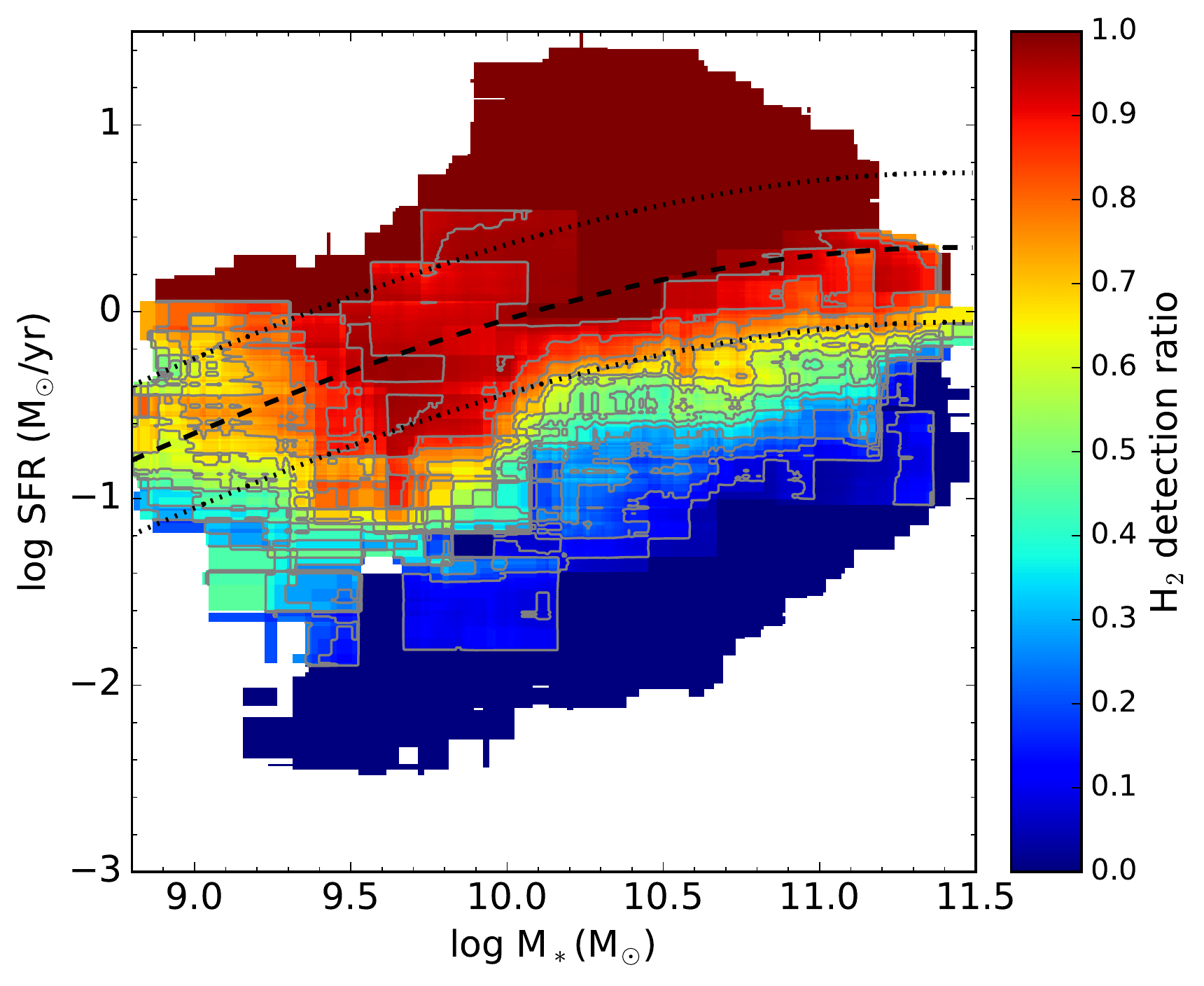}
     \end{center}
 \noindent\textbf{Figure A1.} {\htwo\ detection ratio in xCOLD GASS sample, as a function of stellar mass ($M_{*}$) and star formation rate (SFR), determined within moving boxes of size 0.5 dex in mass and 0.5 dex in SFR. Each galaxy is weighted by a correction factor to account for selection effects in stellar mass. The dashed line indicates the position of the star-forming main sequence defined in \citet{2016MNRAS.462.1749S}. The dotted lines indicate $\pm$0.4 dex scatter around the main sequence, as in \citet{2016MNRAS.462.1749S} and \citet{Saintonge:2017iz}.}
  \label{det_2d}
 \end{figure}

Figure A2 shows \htwo\ detection ratio as a function of SFR (left) and sSFR (right), for galaxies within different stellar mass bins. In the left panel, at log SFR $>$ -0.8 for galaxies with $9.4 <log M_{*} < 10$ (red line) and log SFR $>$ -0.2 for galaxies with log $M_{*} > $ 10 (blue line), the \htwo\ detection ratio is above 80\%, which means the selection effects due to the CO completeness limit should be small. This limit is equivalent to log sSFR $>$ -1.6 Gyr$^{-1}$ for galaxies with log $M_{*} > $ 9.4 as shown in the right panel.

For galaxies below the 80\% detection limit, i.e. in the low SFR regions with rapidly increasing numbers of non-detection in CO, will the results present in the main text be still meaningful? Will they be significantly biased by the increasingly low CO detection ratio? To put it another way, at a given (low) SFR or sSFR, why there are some galaxies with CO detections while others do not?

\defcitealias{Janowiecki:2017bb}{J17}

The SFRs used in our analysis are calculated using the combination of MIR and UV from WISE and GALEX survey, as described in \citeauthor{Janowiecki:2017bb}(2017; hereafter J17). In \citetalias{Janowiecki:2017bb}, SFRs were determined from four cases. The SFRcase\_best flag equals to 1 means from NUV + w4 (22 $\upmu$m); flag = 2 for NUV + w3 (12 $\upmu$m); flag = 6 for NUV or w3 or w4 and flag = 9 for SED SFRs \citep{2011MNRAS.412.1081W} corrected to match UV+MIR. These SFR flags are shown in figure A3, for galaxies with CO detections (left) and CO non-detections (right) on the SFR-$M_*$ plane.

\begin{figure*}[htbp]
     \begin{center}
        \includegraphics[width=180mm]{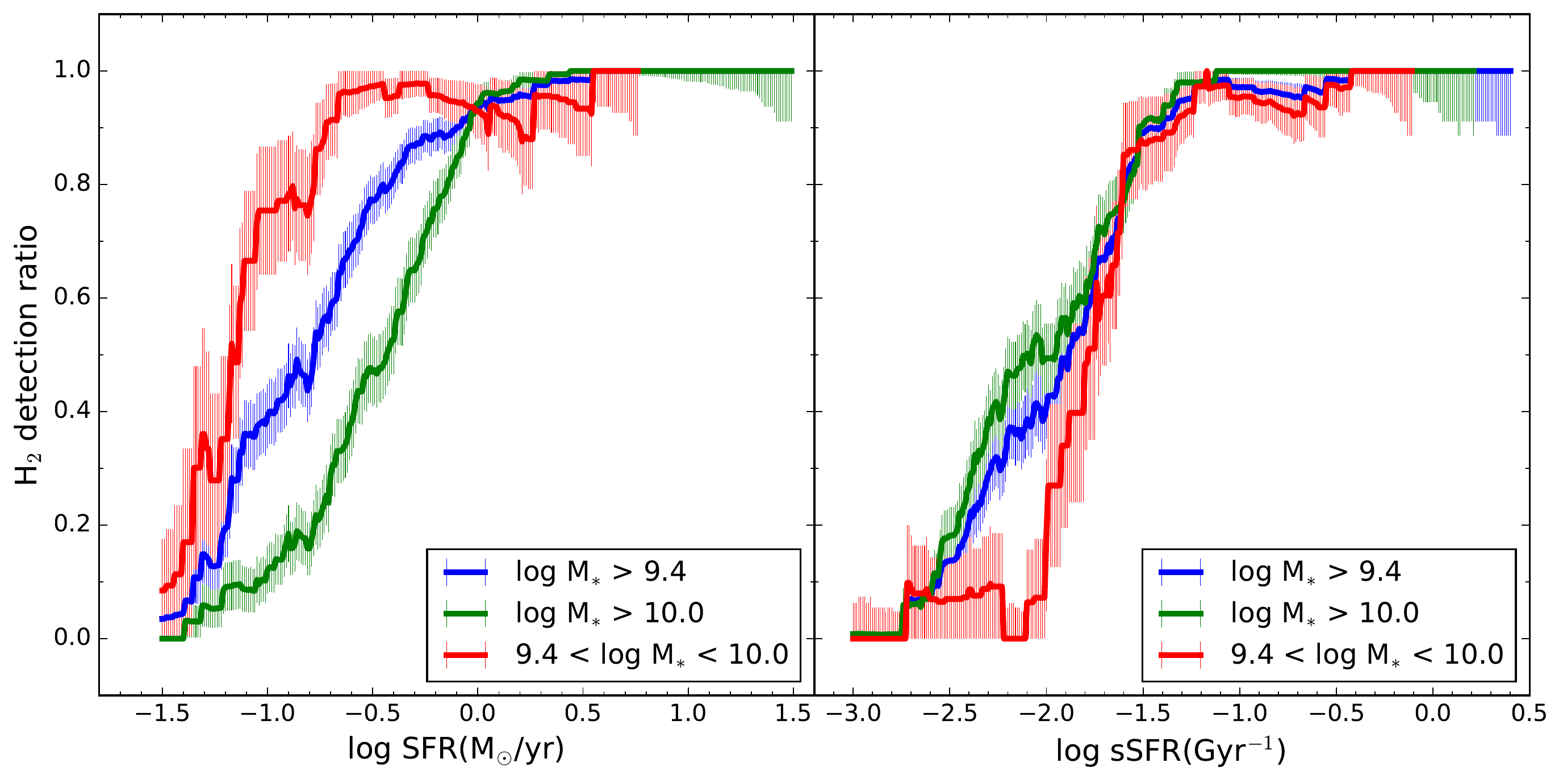}
     \end{center}
 \noindent\textbf{Figure A2.} {\htwo\ detection ratio as a function of SFR (left) and sSFR (right), for galaxies within different stellar mass bins.}
  \label{det_1d}
 \end{figure*}

\begin{figure*}[htbp]
     \begin{center}
        \includegraphics[width=180mm]{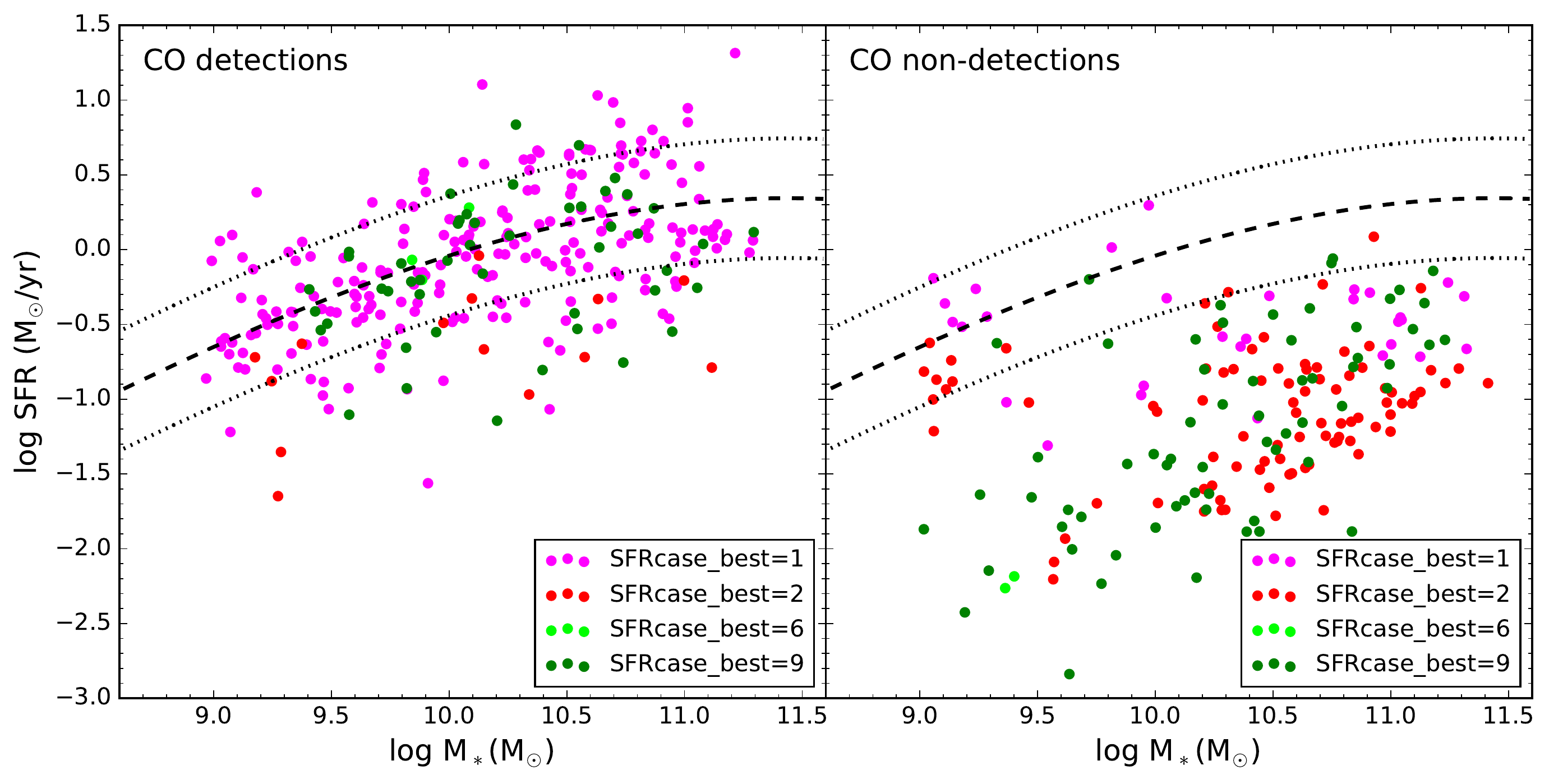}
     \end{center}
 \noindent\textbf{Figure A3.} {The SFR flag in \citet{Janowiecki:2017bb} for galaxies with CO detections (left) and CO non-detections (right) on the SFR-$M_*$ plane. The SFR flag notes how the SFRs were derived: SFRcase\_best = 1 for NUV + w4; 2 for NUV + w3; 6 for NUV or w3 or w4 and 9 for SED SFRs corrected to match UV+MIR.}
  \label{sfr_case}
 \end{figure*}

It is evident that galaxies with CO detections, most of their SFRs were estimated from NUV + w4 (magenta points), some from SED (green points) and only few from NUV + w3 (red points). While for galaxies with CO non-detections, the situation reverses, most of their SFRs were estimated from either NUV + w3 or SED, and only a few from NUV + w4. As discussed in \citetalias{Janowiecki:2017bb}, the w4 emission is a more reliable tracer of the SFR. The w3 emission in reddest galaxies can be entirely attributed to old stellar populations, and not to recent star formation. Comparing the two panels in Figure A3, we suspect the SFRs for galaxies without CO detections, mainly those derived from NUV + w3 or SED, might not be accurate and could be overestimated. Their true SFRs might be lower and fall below the predesigned CO detection limit, and hence are not detected.

We further check the morphology composition of the galaxies with and without CO detections. At log $M_*$ larger than 9.4, galaxies with CO detections have 4.6\% ellipticals, 64.3\% disks and 31.1\% uncertains; galaxies without CO detections have 52\% ellipticals, 10.3\% disks and 37.7\% uncertains. These low SFR ellipticals (whose SFRs are mainly derived from NUV + w3 or SED, hence may not be very accurate) are presumably quenched at high redshifts with little remaining cold gas and ongoing star formation, and may follow different scaling relations than these presented in the current paper. Therefore, if we exclude these old retired elliptical galaxies, the CO detection ratio at the low SFR end will significantly increase, as about 50\% of the galaxies without CO detections are ellipticals. If we select only disk galaxies, the average CO detection ratio will be larger than 80\% at any SFRs, in consistent with the results in \citet{2019ApJ...884L..52Z}.

Putting together, we believe even below the 80\% detection limit, the results presented in the main text might be still meaningful and are not significantly biased by the low CO detection ratio. This explains why the slope and general trend of the various scaling relations remain unchanged towards the low SFR end, even at the lowest SFRs where the detection ratio becomes very low.

On the other hand, the selection effect will affect all scaling relations explored in this paper. Our analysis are primarily comparing the relative differences, for instance, between the $M_{\rm H_2}$-SFR and $\mu$-sSFR relations, not to determine their absolute slope and normalization, hence selection effects are less critical.

\clearpage

\section{The dependence of SFE-sSFR relation on other galaxy properties}

\begin{figure*}[htbp]
     \begin{center}
        \includegraphics[width=180mm]{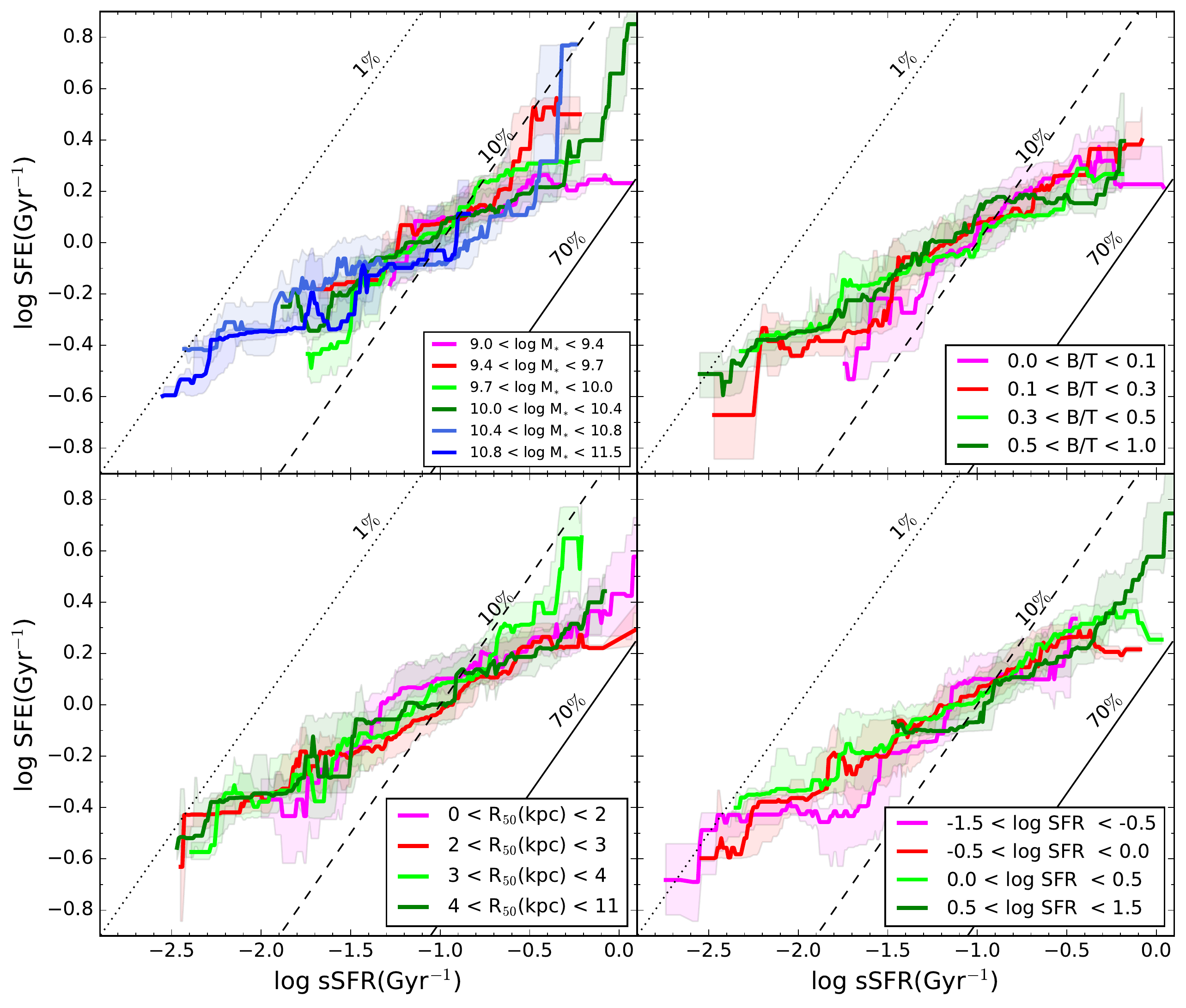}
     \end{center}
 \noindent\textbf{Figure A4.} {As for Figure \ref{u-ssfr}, but for the SFE-sSFR relation. The black diagonal lines are the constant $\mu$ in three different values. Similar to the $\mu$-sSFR relation, the SFE-sSFR relation also shows little or no dependence on other galaxy properties.}
  \label{FIGUREA4}
 \end{figure*}


\clearpage





\begin{thebibliography}{}
\expandafter\ifx\csname natexlab\endcsname\relax\def\natexlab#1{#1}\fi
\providecommand{\url}[1]{\href{#1}{#1}}

\bibitem[{Abazajian {et~al.}(2009)Abazajian, Adelman-McCarthy, Ag{\"u}eros,
  Allam, Prieto, An, Anderson, Anderson, Annis, Bahcall, Bailer-Jones,
  Barentine, Bassett, Becker, Beers, Bell, Belokurov, Berlind, Berman,
  Bernardi, Bickerton, Bizyaev, Blakeslee, Blanton, Bochanski, Boroski,
  Brewington, Brinchmann, Brinkmann, Brunner, Budav{\'a}ri, Carey, Carliles,
  Carr, Castander, Cinabro, Connolly, Csabai, Cunha, Czarapata, Davenport,
  de~Haas, Dilday, Doi, Eisenstein, Evans, Evans, Fan, Friedman, Frieman,
  Fukugita, G{\"a}nsicke, Gates, Gillespie, Gilmore, Gonzalez, Gonzalez,
  Grebel, Gunn, Gy{\"o}ry, Hall, Harding, Harris, Harvanek, Hawley, Hayes,
  Heckman, Hendry, Hennessy, Hindsley, Hoblitt, Hogan, Hogg, Holtzman, Hyde,
  Ichikawa, Ichikawa, Im, Ivezi{\'c}, Jester, Jiang, Johnson, Jorgensen,
  Juri{\'c}, Kent, Kessler, Kleinman, Knapp, Konishi, Kron, Krzesinski,
  Kuropatkin, Lampeitl, Lebedeva, Lee, Lee, Leger, L{\'e}pine, Li, Lima, Lin,
  Long, Loomis, Loveday, Lupton, Magnier, Malanushenko, Malanushenko,
  Mandelbaum, Margon, Marriner, Mart{\'\i}nez-Delgado, Matsubara, McGehee,
  McKay, Meiksin, Morrison, Mullally, Munn, Murphy, Nash, Nebot, Neilsen,
  Newberg, Newman, Nichol, Nicinski, Nieto-Santisteban, Nitta, Okamura,
  Oravetz, Ostriker, Owen, Padmanabhan, Pan, Park, Pauls, Peoples, Percival,
  Pier, Pope, Pourbaix, Price, Purger, Quinn, Raddick, Fiorentin, Richards,
  Richmond, Riess, Rix, Rockosi, Sako, Schlegel, Schneider, Scholz, Schreiber,
  Schwope, Seljak, Sesar, Sheldon, Shimasaku, Sibley, Simmons, Sivarani, Smith,
  Smith, Smol{\v c}i{\'c}, Snedden, Stebbins, Steinmetz, Stoughton, Strauss,
  SubbaRao, Suto, Szalay, Szapudi, Szkody, Tanaka, Tegmark, Teodoro, Thakar,
  Tremonti, Tucker, Uomoto, Vanden~Berk, Vandenberg, Vidrih, Vogeley, Voges,
  Vogt, Wadadekar, Watters, Weinberg, West, White, Wilhite, Wonders, Yanny,
  Yocum, York, Zehavi, Zibetti, \& Zucker}]{Abazajian:2009ef}
Abazajian, K.~N., Adelman-McCarthy, J.~K., Ag{\"u}eros, M.~A., {et~al.} 2009,
  The Astrophysical Journal Supplement Series, 182, 543

\bibitem[{Accurso {et~al.}(2017)Accurso, Saintonge, Catinella, Cortese,
  Dav{\'e}, Dunsheath, Genzel, Gracia-Carpio, Heckman, Jimmy, Kramer, Li, Lutz,
  Schiminovich, Schuster, Sternberg, Sturm, Tacconi, Tran, \&
  Wang}]{Accurso:2017kf}
Accurso, G., Saintonge, A., Catinella, B., {et~al.} 2017, Monthly Notices of
  the Royal Astronomical Society, 1

\bibitem[{Aravena {et~al.}(2019)Aravena, Decarli, Gónzalez-López, Boogaard,
  Walter, Carilli, Popping, Weiss, Assef, Bacon, \& et~al.}]{Aravena_2019}
Aravena, M., Decarli, R., Gónzalez-López, J., {et~al.} 2019, The
  Astrophysical Journal, 882, 136

\bibitem[{{Barrera-Ballesteros}(2019)}]{2019arXiv191207633B}
{Barrera-Ballesteros}, J.~K. 2019, arXiv e-prints, arXiv:1912.07633

\bibitem[{{Barrera-Ballesteros} {et~al.}(2018){Barrera-Ballesteros}, {Heckman},
  {S{\'a}nchez}, {Zakamska}, {Cleary}, {Zhu}, {Brinkmann}, {Drory}, \& {THE
  MaNGA TEAM}}]{2018ApJ...852...74B}
{Barrera-Ballesteros}, J.~K., {Heckman}, T., {S{\'a}nchez}, S.~F., {et~al.}
  2018, \apj, 852, 74

\bibitem[{{Barrera-Ballesteros} {et~al.}(2020){Barrera-Ballesteros}, {Utomo},
  {Bolatto}, {S{\'a}nchez}, {Vogel}, {Wong}, {Levy}, {Colombo}, {Kalinova},
  {Teuben}, {Garc{\'\i}a-Benito}, {Husemann}, {Mast}, \&
  {Blitz}}]{2020MNRAS.492.2651B}
{Barrera-Ballesteros}, J.~K., {Utomo}, D., {Bolatto}, A.~D., {et~al.} 2020,
  \mnras, 492, 2651

\bibitem[{{Bigiel} {et~al.}(2008){Bigiel}, {Leroy}, {Walter}, {Brinks}, {de
  Blok}, {Madore}, \& {Thornley}}]{2008AJ....136.2846B}
{Bigiel}, F., {Leroy}, A., {Walter}, F., {et~al.} 2008, \aj, 136, 2846

\bibitem[{{Bolatto} {et~al.}(2017){Bolatto}, {Wong}, {Utomo}, {Blitz}, {Vogel},
  {S{\'a}nchez}, {Barrera-Ballesteros}, {Cao}, {Colombo}, {Dannerbauer},
  {Garc{\'\i}a-Benito}, {Herrera-Camus}, {Husemann}, {Kalinova}, {Leroy},
  {Leung}, {Levy}, {Mast}, {Ostriker}, {Rosolowsky}, {Sandstrom}, {Teuben},
  {van de Ven}, \& {Walter}}]{2017ApJ...846..159B}
{Bolatto}, A.~D., {Wong}, T., {Utomo}, D., {et~al.} 2017, \apj, 846, 159

\bibitem[{{Bothwell} {et~al.}(2014){Bothwell}, {Wagg}, {Cicone}, {Maiolino},
  {M{\o}ller}, {Aravena}, {De Breuck}, {Peng}, {Espada}, {Hodge},
  {Impellizzeri}, {Mart{\'\i}n}, {Riechers}, \& {Walter}}]{2014MNRAS.445.2599B}
{Bothwell}, M.~S., {Wagg}, J., {Cicone}, C., {et~al.} 2014, \mnras, 445, 2599

\bibitem[{{Bouch{\'e}} {et~al.}(2007){Bouch{\'e}}, {Cresci}, {Davies},
  {Eisenhauer}, {F{\"o}rster Schreiber}, {Genzel}, {Gillessen}, {Lehnert},
  {Lutz}, {Nesvadba}, {Shapiro}, {Sternberg}, {Tacconi}, {Verma}, {Cimatti},
  {Daddi}, {Renzini}, {Erb}, {Shapley}, \& {Steidel}}]{2007ApJ...671..303B}
{Bouch{\'e}}, N., {Cresci}, G., {Davies}, R., {et~al.} 2007, \apj, 671, 303

\bibitem[{{Brinchmann} {et~al.}(2004){Brinchmann}, {Charlot}, {White},
  {Tremonti}, {Kauffmann}, {Heckman}, \& {Brinkmann}}]{2004MNRAS.351.1151B}
{Brinchmann}, J., {Charlot}, S., {White}, S.~D.~M., {et~al.} 2004, \mnras, 351,
  1151

\bibitem[{{Catinella} {et~al.}(2010){Catinella}, {Schiminovich}, {Kauffmann},
  {Fabello}, {Wang}, {Hummels}, {Lemonias}, {Moran}, {Wu}, {Giovanelli},
  {Haynes}, {Heckman}, {Basu-Zych}, {Blanton}, {Brinchmann}, {Budav{\'a}ri},
  {Gon{\c{c}}alves}, {Johnson}, {Kennicutt}, {Madore}, {Martin}, {Rich},
  {Tacconi}, {Thilker}, {Wild}, \& {Wyder}}]{2010MNRAS.403..683C}
{Catinella}, B., {Schiminovich}, D., {Kauffmann}, G., {et~al.} 2010, \mnras,
  403, 683

\bibitem[{{Chabrier}(2003)}]{2003PASP..115..763C}
{Chabrier}, G. 2003, \pasp, 115, 763

\bibitem[{Cibinel {et~al.}(2019)Cibinel, Daddi, Sargent, Le~Floc’h, Liu,
  Bournaud, Oesch, Amram, Calabrò, Duc, \& et~al.}]{Cibinel_2019}
Cibinel, A., Daddi, E., Sargent, M.~T., {et~al.} 2019, Monthly Notices of the
  Royal Astronomical Society, 485, 5631–5651

\bibitem[{{Cicone} {et~al.}(2017){Cicone}, {Bothwell}, {Wagg}, {M{\o}ller}, {De
  Breuck}, {Zhang}, {Mart{\'\i}n}, {Maiolino}, {Severgnini}, {Aravena},
  {Belfiore}, {Espada}, {Fl{\"u}tsch}, {Impellizzeri}, {Peng}, {Raj},
  {Ram{\'\i}rez-Olivencia}, {Riechers}, \& {Schawinski}}]{2017A&A...604A..53C}
{Cicone}, C., {Bothwell}, M., {Wagg}, J., {et~al.} 2017, \aap, 604, A53

\bibitem[{{Conselice}(2014)}]{2014ARA&A..52..291C}
{Conselice}, C.~J. 2014, \araa, 52, 291

\bibitem[{{Coogan} {et~al.}(2018){Coogan}, {Daddi}, {Sargent}, {Strazzullo},
  {Valentino}, {Gobat}, {Magdis}, {Bethermin}, {Pannella}, {Onodera}, {Liu},
  {Cimatti}, {Dannerbauer}, {Carollo}, {Renzini}, \&
  {Tremou}}]{2018MNRAS.479..703C}
{Coogan}, R.~T., {Daddi}, E., {Sargent}, M.~T., {et~al.} 2018, \mnras, 479, 703

\bibitem[{{Daddi} {et~al.}(2007){Daddi}, {Dickinson}, {Morrison}, {Chary},
  {Cimatti}, {Elbaz}, {Frayer}, {Renzini}, {Pope}, {Alexander}, {Bauer},
  {Giavalisco}, {Huynh}, {Kurk}, \& {Mignoli}}]{2007ApJ...670..156D}
{Daddi}, E., {Dickinson}, M., {Morrison}, G., {et~al.} 2007, \apj, 670, 156


\bibitem[{{Daddi} {et~al.}(2010){Daddi}, {Elbaz}, {Walter}, {Bournaud},
  {Salmi}, {Carilli}, {Dannerbauer}, {Dickinson}, {Monaco}, \&
  {Riechers}}]{2010ApJ...714L.118D}
{Daddi}, E., {Elbaz}, D., {Walter}, F., {et~al.} 2010, \apjl, 714, L118

\bibitem[{{de los Reyes} \& {Kennicutt}(2019)}]{2019ApJ...872...16D}
{de los Reyes}, M. A.~C., \& {Kennicutt}, Robert~C., J. 2019, \apj, 872, 16

\bibitem[{{Dekel} \& {Mandelker}(2014)}]{2014MNRAS.444.2071D}
{Dekel}, A., \& {Mandelker}, N. 2014, \mnras, 444, 2071

\bibitem[{{Dey} {et~al.}(2019){Dey}, {Rosolowsky}, {Cao}, {Bolatto}, {Sanchez},
  {Utomo}, {Colombo}, {Kalinova}, {Wong}, {Blitz}, {Vogel}, {Loeppky}, \&
  {Garc{\'\i}a-Benito}}]{2019MNRAS.488.1926D}
{Dey}, B., {Rosolowsky}, E., {Cao}, Y., {et~al.} 2019, \mnras, 488, 1926

\bibitem[{{Dopita}(1985)}]{1985ApJ...295L...5D}
{Dopita}, M.~A. 1985, \apjl, 295, L5

\bibitem[{{Dopita} \& {Ryder}(1994)}]{1994ApJ...430..163D}
{Dopita}, M.~A., \& {Ryder}, S.~D. 1994, \apj, 430, 163

\bibitem[{Duncan {et~al.}(2019)Duncan, Conselice, Mundy, Bell, Donley,
  Galametz, Guo, Grogin, Hathi, Kartaltepe, \& et~al.}]{Duncan_2019}
Duncan, K., Conselice, C.~J., Mundy, C., {et~al.} 2019, The Astrophysical
  Journal, 876, 110

\bibitem[{{Elbaz} {et~al.}(2007){Elbaz}, {Daddi}, {Le Borgne}, {Dickinson},
  {Alexander}, {Chary}, {Starck}, {Brand t}, {Kitzbichler}, {MacDonald},
  {Nonino}, {Popesso}, {Stern}, \& {Vanzella}}]{2007A&A...468...33E}
{Elbaz}, D., {Daddi}, E., {Le Borgne}, D., {et~al.} 2007, \aap, 468, 33

\bibitem[{{Elbaz} {et~al.}(2011){Elbaz}, {Dickinson}, {Hwang},
  {D{\'\i}az-Santos}, {Magdis}, {Magnelli}, {Le Borgne}, {Galliano},
  {Pannella}, {Chanial}, {Armus}, {Charmandaris}, {Daddi}, {Aussel}, {Popesso},
  {Kartaltepe}, {Altieri}, {Valtchanov}, {Coia}, {Dannerbauer}, {Dasyra},
  {Leiton}, {Mazzarella}, {Alexander}, {Buat}, {Burgarella}, {Chary}, {Gilli},
  {Ivison}, {Juneau}, {Le Floc'h}, {Lutz}, {Morrison}, {Mullaney}, {Murphy},
  {Pope}, {Scott}, {Brodwin}, {Calzetti}, {Cesarsky}, {Charlot}, {Dole},
  {Eisenhardt}, {Ferguson}, {F{\"o}rster Schreiber}, {Frayer}, {Giavalisco},
  {Huynh}, {Koekemoer}, {Papovich}, {Reddy}, {Surace}, {Teplitz}, {Yun}, \&
  {Wilson}}]{2011A&A...533A.119E}
{Elbaz}, D., {Dickinson}, M., {Hwang}, H.~S., {et~al.} 2011, \aap, 533, A119

\bibitem[{{Ellison} {et~al.}(2020{\natexlab{a}}){Ellison}, {Thorp}, {Pan},
  {Lin}, {Scudder}, {Bluck}, {S{\'a}nchez}, \& {Sargent}}]{2020MNRAS.492.6027E}
{Ellison}, S.~L., {Thorp}, M.~D., {Pan}, H.-A., {et~al.} 2020{\natexlab{a}},
  \mnras, 492, 6027

\bibitem[{{Ellison} {et~al.}(2020{\natexlab{b}}){Ellison}, {Thorp}, {Lin},
  {Pan}, {Bluck}, {Scudder}, {Teimoorinia}, {S{\'a}nchez}, \&
  {Sargent}}]{2020MNRAS.493L..39E}
{Ellison}, S.~L., {Thorp}, M.~D., {Lin}, L., {et~al.} 2020{\natexlab{b}},
  \mnras, 493, L39

\bibitem[{{Elmegreen} \& {Efremov}(1997)}]{1997ApJ...480..235E}
{Elmegreen}, B.~G., \& {Efremov}, Y.~N. 1997, \apj, 480, 235

\bibitem[{Ferreira {et~al.}(2020)Ferreira, Conselice, Duncan, Cheng, Griffiths,
  \& Whitney}]{Ferreira_2020}
Ferreira, L., Conselice, C.~J., Duncan, K., {et~al.} 2020, The Astrophysical
  Journal, 895, 115

\bibitem[{{Franx} {et~al.}(2008){Franx}, {van Dokkum}, {F{\"o}rster Schreiber},
  {Wuyts}, {Labb{\'e}}, \& {Toft}}]{2008ApJ...688..770F}
{Franx}, M., {van Dokkum}, P.~G., {F{\"o}rster Schreiber}, N.~M., {et~al.}
  2008, \apj, 688, 770

\bibitem[{{Freundlich} {et~al.}(2019){Freundlich}, {Combes}, {Tacconi},
  {Genzel}, {Garcia-Burillo}, {Neri}, {Contini}, {Bolatto}, {Lilly},
  {Salom{\'e}}, {Bicalho}, {Boissier}, {Boone}, {Bouch{\'e}}, {Bournaud},
  {Burkert}, {Carollo}, {Cooper}, {Cox}, {Feruglio}, {F{\"o}rster Schreiber},
  {Juneau}, {Lippa}, {Lutz}, {Naab}, {Renzini}, {Saintonge}, {Sternberg},
  {Walter}, {Weiner}, {Wei{\ss}}, \& {Wuyts}}]{2019A&A...622A.105F}
{Freundlich}, J., {Combes}, F., {Tacconi}, L.~J., {et~al.} 2019, \aap, 622,
  A105

\bibitem[{{Gao} \& {Solomon}(2004)}]{2004ApJ...606..271G}
{Gao}, Y., \& {Solomon}, P.~M. 2004, \apj, 606, 271

\bibitem[{{Gensior} {et~al.}(2020){Gensior}, {Kruijssen}, \&
  {Keller}}]{2020MNRAS.495..199G}
{Gensior}, J., {Kruijssen}, J.~M.~D., \& {Keller}, B.~W. 2020, \mnras, 495, 199

\bibitem[{{Genzel} {et~al.}(2010){Genzel}, {Tacconi}, {Gracia-Carpio},
  {Sternberg}, {Cooper}, {Shapiro}, {Bolatto}, {Bouch{\'e}}, {Bournaud},
  {Burkert}, {Combes}, {Comerford}, {Cox}, {Davis}, {Schreiber},
  {Garcia-Burillo}, {Lutz}, {Naab}, {Neri}, {Omont}, {Shapley}, \&
  {Weiner}}]{2010MNRAS.407.2091G}
{Genzel}, R., {Tacconi}, L.~J., {Gracia-Carpio}, J., {et~al.} 2010, \mnras,
  407, 2091

\bibitem[{{Genzel} {et~al.}(2012){Genzel}, {Tacconi}, {Combes}, {Bolatto},
  {Neri}, {Sternberg}, {Cooper}, {Bouch{\'e}}, {Bournaud}, {Burkert},
  {Comerford}, {Cox}, {Davis}, {F{\"o}rster Schreiber}, {Garcia-Burillo},
  {Gracia-Carpio}, {Lutz}, {Naab}, {Newman}, {Saintonge}, {Shapiro}, {Shapley},
  \& {Weiner}}]{2012ApJ...746...69G}
{Genzel}, R., {Tacconi}, L.~J., {Combes}, F., {et~al.} 2012, \apj, 746, 69

\bibitem[{{Genzel} {et~al.}(2014){Genzel}, {F{\"o}rster Schreiber}, {Lang},
  {Tacchella}, {Tacconi}, {Wuyts}, {Bandara}, {Burkert}, {Buschkamp},
  {Carollo}, {Cresci}, {Davies}, {Eisenhauer}, {Hicks}, {Kurk}, {Lilly},
  {Lutz}, {Mancini}, {Naab}, {Newman}, {Peng}, {Renzini}, {Shapiro Griffin},
  {Sternberg}, {Vergani}, {Wisnioski}, {Wuyts}, \&
  {Zamorani}}]{2014ApJ...785...75G}
{Genzel}, R., {F{\"o}rster Schreiber}, N.~M., {Lang}, P., {et~al.} 2014, \apj,
  785, 75

\bibitem[{Genzel {et~al.}(2015)Genzel, Tacconi, Lutz, Saintonge, Berta,
  Magnelli, Combes, Garc{\'\i}a-Burillo, Neri, Bolatto, Contini, Lilly,
  Boissier, Boone, Bouch{\'e}, Bournaud, Burkert, Carollo, Colina, Cooper, Cox,
  Feruglio, F{\"o}rster~Schreiber, Freundlich, Gracia-Carpio, Juneau, Kovac,
  Lippa, Naab, Salome, Renzini, Sternberg, Walter, Weiner, Weiss, \&
  Wuyts}]{Genzel:2015fq}
Genzel, R., Tacconi, L.~J., Lutz, D., {et~al.} 2015, The Astrophysical Journal,
  800, 20

\bibitem[{{Grossi} {et~al.}(2016){Grossi}, {Corbelli}, {Bizzocchi},
  {Giovanardi}, {Bomans}, {Coelho}, {De Looze}, {Gon{\c{c}}alves}, {Hunt},
  {Leonardo}, {Madden}, {Men{\'e}ndez-Delmestre}, {Pappalardo}, \&
  {Riguccini}}]{2016A&A...590A..27G}
{Grossi}, M., {Corbelli}, E., {Bizzocchi}, L., {et~al.} 2016, \aap, 590, A27

\bibitem[{Haynes {et~al.}(2011)Haynes, Giovanelli, Martin, Hess, Saintonge,
  Adams, Hallenbeck, Hoffman, Huang, Kent, Koopmann, Papastergis, Stierwalt,
  Balonek, Craig, Higdon, Kornreich, Miller, O'Donoghue, Olowin, Rosenberg,
  Spekkens, Troischt, \& Wilcots}]{Haynes:2011en}
Haynes, M.~P., Giovanelli, R., Martin, A.~M., {et~al.} 2011, The Astronomical
  Journal, 142, 170

\bibitem[{Huang \& Kauffmann(2014)}]{Huang:2014ko}
Huang, M.-L., \& Kauffmann, G. 2014, Monthly Notices of the Royal Astronomical
  Society, 443, 1329

\bibitem[{Janowiecki {et~al.}(2017)Janowiecki, Catinella, Cortese, Saintonge,
  Brown, \& Wang}]{Janowiecki:2017bb}
Janowiecki, S., Catinella, B., Cortese, L., {et~al.} 2017, Monthly Notices of
  the Royal Astronomical Society, stx046

\bibitem[{{Kauffmann} {et~al.}(1993){Kauffmann}, {White}, \&
  {Guiderdoni}}]{1993MNRAS.264..201K}
{Kauffmann}, G., {White}, S.~D.~M., \& {Guiderdoni}, B. 1993, \mnras, 264, 201

\bibitem[{{Kennicutt}(1998)}]{1998ApJ...498..541K}
{Kennicutt}, Robert~C., J. 1998, \apj, 498, 541

\bibitem[{{Kennicutt} \& {Evans}(2012)}]{2012ARA&A..50..531K}
{Kennicutt}, R.~C., \& {Evans}, N.~J. 2012, \araa, 50, 531

\bibitem[{{Kreckel} {et~al.}(2018){Kreckel}, {Faesi}, {Kruijssen}, {Schruba},
  {Groves}, {Leroy}, {Bigiel}, {Blanc}, {Chevance}, {Herrera}, {Hughes},
  {McElroy}, {Pety}, {Querejeta}, {Rosolowsky}, {Schinnerer}, {Sun}, {Usero},
  \& {Utomo}}]{2018ApJ...863L..21K}
{Kreckel}, K., {Faesi}, C., {Kruijssen}, J.~M.~D., {et~al.} 2018, \apjl, 863,
  L21

\bibitem[{{Krumholz} \& {McKee}(2005)}]{2005ApJ...630..250K}
{Krumholz}, M.~R., \& {McKee}, C.~F. 2005, \apj, 630, 250

\bibitem[{Leroy {et~al.}(2008)Leroy, Walter, Brinks, Bigiel, de~Blok, Madore,
  \& Thornley}]{Leroy:2008jk}
Leroy, A.~K., Walter, F., Brinks, E., {et~al.} 2008, The Astronomical Journal,
  136, 2782

\bibitem[{{Leroy} {et~al.}(2013){Leroy}, {Walter}, {Sandstrom}, {Schruba},
  {Munoz-Mateos}, {Bigiel}, {Bolatto}, {Brinks}, {de Blok}, {Meidt}, {Rix},
  {Rosolowsky}, {Schinnerer}, {Schuster}, \& {Usero}}]{2013AJ....146...19L}
{Leroy}, A.~K., {Walter}, F., {Sandstrom}, K., {et~al.} 2013, \aj, 146, 19

\bibitem[{Lilly {et~al.}(2013)Lilly, Carollo, Pipino, Renzini, \&
  Peng}]{Lilly:2013ko}
Lilly, S.~J., Carollo, C.~M., Pipino, A., Renzini, A., \& Peng, Y. 2013, The
  Astrophysical Journal, 772, 119

\bibitem[{{Lilly} {et~al.}(1996){Lilly}, {Le Fevre}, {Hammer}, \&
  {Crampton}}]{1996ApJ...460L...1L}
{Lilly}, S.~J., {Le Fevre}, O., {Hammer}, F., \& {Crampton}, D. 1996, \apjl,
  460, L1

\bibitem[{{Lin} {et~al.}(2019){Lin}, {Pan}, {Ellison}, {Belfiore}, {Shi},
  {S{\'a}nchez}, {Hsieh}, {Rowland s}, {Ramya}, {Thorp}, {Li}, \&
  {Maiolino}}]{2019ApJ...884L..33L}
{Lin}, L., {Pan}, H.-A., {Ellison}, S.~L., {et~al.} 2019, \apjl, 884, L33

\bibitem[{{Liu} {et~al.}(2015){Liu}, {Gao}, \& {Greve}}]{2015ApJ...805...31L}
{Liu}, L., {Gao}, Y., \& {Greve}, T.~R. 2015, \apj, 805, 31

\bibitem[{{Madau} \& {Dickinson}(2014)}]{2014ARA&A..52..415M}
{Madau}, P., \& {Dickinson}, M. 2014, \araa, 52, 415

\bibitem[{{Maiolino} \& {Mannucci}(2019)}]{2019A&ARv..27....3M}
{Maiolino}, R., \& {Mannucci}, F. 2019, \aapr, 27, 3

\bibitem[{{Martig} {et~al.}(2009){Martig}, {Bournaud}, {Teyssier}, \&
  {Dekel}}]{2009ApJ...707..250M}
{Martig}, M., {Bournaud}, F., {Teyssier}, R., \& {Dekel}, A. 2009, \apj, 707,
  250

\bibitem[{{Mendel} {et~al.}(2014){Mendel}, {Simard}, {Palmer}, {Ellison}, \&
  {Patton}}]{2014ApJS..210....3M}
{Mendel}, J.~T., {Simard}, L., {Palmer}, M., {Ellison}, S.~L., \& {Patton},
  D.~R. 2014, \apjs, 210, 3

\bibitem[{{Mo} {et~al.}(1998){Mo}, {Mao}, \& {White}}]{1998MNRAS.295..319M}
{Mo}, H.~J., {Mao}, S., \& {White}, S. D.~M. 1998, \mnras, 295, 319

\bibitem[{{Morselli} {et~al.}(2020){Morselli}, {Rodighiero}, {Enia},
  {Corbelli}, {Casasola}, {Rodr{\'\i}guez-Mu{\~n}oz}, {Renzini}, {Tacchella},
  {Baronchelli}, {Bianchi}, {Cassata}, {Franceschini}, {Mancini}, {Negrello},
  {Popesso}, \& {Romano}}]{2020MNRAS.496.4606M}
{Morselli}, L., {Rodighiero}, G., {Enia}, A., {et~al.} 2020, \mnras, 496, 4606

\bibitem[{{Murray}(2011)}]{2011ApJ...729..133M}
{Murray}, N. 2011, \apj, 729, 133

\bibitem[{{Navarro} \& {Steinmetz}(2000)}]{2000ApJ...538..477N}
{Navarro}, J.~F., \& {Steinmetz}, M. 2000, \apj, 538, 477

\bibitem[{{Noeske} {et~al.}(2007){Noeske}, {Weiner}, {Faber}, {Papovich},
  {Koo}, {Somerville}, {Bundy}, {Conselice}, {Newman}, {Schiminovich}, {Le
  Floc'h}, {Coil}, {Rieke}, {Lotz}, {Primack}, {Barmby}, {Cooper}, {Davis},
  {Ellis}, {Fazio}, {Guhathakurta}, {Huang}, {Kassin}, {Martin}, {Phillips},
  {Rich}, {Small}, {Willmer}, \& {Wilson}}]{2007ApJ...660L..43N}
{Noeske}, K.~G., {Weiner}, B.~J., {Faber}, S.~M., {et~al.} 2007, \apjl, 660,
  L43

\bibitem[{{Papovich} {et~al.}(2016){Papovich}, {Labb{\'e}}, {Glazebrook},
  {Quadri}, {Bekiaris}, {Dickinson}, {Finkelstein}, {Fisher}, {Inami},
  {Livermore}, {Spitler}, {Straatman}, \& {Tran}}]{2016NatAs...1E...3P}
{Papovich}, C., {Labb{\'e}}, I., {Glazebrook}, K., {et~al.} 2016, Nature
  Astronomy, 1, 0003

\bibitem[{{Peebles}(1969)}]{1969ApJ...155..393P}
{Peebles}, P.~J.~E. 1969, \apj, 155, 393

\bibitem[{Peng {et~al.}(2015)Peng, Maiolino, \& Cochrane}]{Peng:2015bq}
Peng, Y., Maiolino, R., \& Cochrane, R. 2015, Nature, 521, 192

\bibitem[{Peng \& Maiolino(2014)}]{Peng:2014hn}
Peng, Y.~j., \& Maiolino, R. 2014, Monthly Notices of the Royal Astronomical
  Society, 443, 3643

\bibitem[{{Peng} \& {Renzini}(2020)}]{2020MNRAS.491L..51P}
{Peng}, Y.-j., \& {Renzini}, A. 2020, \mnras, 491, L51

\bibitem[{{Peng} {et~al.}(2010){Peng}, {Lilly}, {Kova{\v{c}}}, {Bolzonella},
  {Pozzetti}, {Renzini}, {Zamorani}, {Ilbert}, {Knobel}, {Iovino}, {Maier},
  {Cucciati}, {Tasca}, {Carollo}, {Silverman}, {Kampczyk}, {de Ravel},
  {Sanders}, {Scoville}, {Contini}, {Mainieri}, {Scodeggio}, {Kneib}, {Le
  F{\`e}vre}, {Bardelli}, {Bongiorno}, {Caputi}, {Coppa}, {de la Torre},
  {Franzetti}, {Garilli}, {Lamareille}, {Le Borgne}, {Le Brun}, {Mignoli},
  {Perez Montero}, {Pello}, {Ricciardelli}, {Tanaka}, {Tresse}, {Vergani},
  {Welikala}, {Zucca}, {Oesch}, {Abbas}, {Barnes}, {Bordoloi}, {Bottini},
  {Cappi}, {Cassata}, {Cimatti}, {Fumana}, {Hasinger}, {Koekemoer},
  {Leauthaud}, {Maccagni}, {Marinoni}, {McCracken}, {Memeo}, {Meneux}, {Nair},
  {Porciani}, {Presotto}, \& {Scaramella}}]{2010ApJ...721..193P}
{Peng}, Y.-j., {Lilly}, S.~J., {Kova{\v{c}}}, K., {et~al.} 2010, \apj, 721, 193


\bibitem[{{Piotrowska} {et~al.}(2020){Piotrowska}, {Bluck}, {Maiolino},
  {Concas}, \& {Peng}}]{2020MNRAS.492L...6P}
{Piotrowska}, J.~M., {Bluck}, A. F.~L., {Maiolino}, R., {Concas}, A., \&
  {Peng}, Y. 2020, \mnras, 492, L6

\bibitem[{{Popesso} {et~al.}(2019{\natexlab{a}}){Popesso}, {Concas},
  {Morselli}, {Schreiber}, {Rodighiero}, {Cresci}, {Belli}, {Erfanianfar},
  {Mancini}, {Inami}, {Dickinson}, {Ilbert}, {Pannella}, \&
  {Elbaz}}]{2019MNRAS.483.3213P}
{Popesso}, P., {Concas}, A., {Morselli}, L., {et~al.} 2019{\natexlab{a}},
  \mnras, 483, 3213

\bibitem[{{Popesso} {et~al.}(2019{\natexlab{b}}){Popesso}, {Morselli},
  {Concas}, {Schreiber}, {Rodighiero}, {Cresci}, {Belli}, {Ilbert},
  {Erfanianfar}, {Mancini}, {Inami}, {Dickinson}, {Pannella}, \&
  {Elbaz}}]{2019MNRAS.490.5285P}
{Popesso}, P., {Morselli}, L., {Concas}, A., {et~al.} 2019{\natexlab{b}},
  \mnras, 490, 5285

\bibitem[{{Renzini}(2016)}]{2016MNRAS.460L..45R}
{Renzini}, A. 2016, \mnras, 460, L45

\bibitem[{{Renzini}(2020)}]{2020MNRAS.495L..42R}
{Renzini}, A. 2020, \mnras, 495, L42

\bibitem[{{Renzini} \& {Peng}(2015)}]{2015ApJ...801L..29R}
{Renzini}, A., \& {Peng}, Y.-j. 2015, \apjl, 801, L29

\bibitem[{{Rodighiero} {et~al.}(2015){Rodighiero}, {Brusa}, {Daddi},
  {Negrello}, {Mullaney}, {Delvecchio}, {Lutz}, {Renzini}, {Franceschini},
  {Baronchelli}, {Pozzi}, {Gruppioni}, {Strazzullo}, {Cimatti}, \&
  {Silverman}}]{2015ApJ...800L..10R}
{Rodighiero}, G., {Brusa}, M., {Daddi}, E., {et~al.} 2015, \apjl, 800, L10

\bibitem[{{Saintonge} {et~al.}(2011{\natexlab{a}}){Saintonge}, {Kauffmann},
  {Wang}, {Kramer}, {Tacconi}, {Buchbender}, {Catinella}, {Graci{\'a}-Carpio},
  {Cortese}, {Fabello}, {Fu}, {Genzel}, {Giovanelli}, {Guo}, {Haynes},
  {Heckman}, {Krumholz}, {Lemonias}, {Li}, {Moran}, {Rodriguez-Fernand ez},
  {Schiminovich}, {Schuster}, \& {Sievers}}]{2011MNRAS.415...61S}
{Saintonge}, A., {Kauffmann}, G., {Wang}, J., {et~al.} 2011{\natexlab{a}},
  \mnras, 415, 61

\bibitem[{{Saintonge} {et~al.}(2011{\natexlab{b}}){Saintonge}, {Kauffmann},
  {Kramer}, {Tacconi}, {Buchbender}, {Catinella}, {Fabello},
  {Graci{\'a}-Carpio}, {Wang}, {Cortese}, {Fu}, {Genzel}, {Giovanelli}, {Guo},
  {Haynes}, {Heckman}, {Krumholz}, {Lemonias}, {Li}, {Moran},
  {Rodriguez-Fernandez}, {Schiminovich}, {Schuster}, \&
  {Sievers}}]{2011MNRAS.415...32S}
{Saintonge}, A., {Kauffmann}, G., {Kramer}, C., {et~al.} 2011{\natexlab{b}},
  \mnras, 415, 32

\bibitem[{{Saintonge} {et~al.}(2012){Saintonge}, {Tacconi}, {Fabello}, {Wang},
  {Catinella}, {Genzel}, {Graci{\'a}-Carpio}, {Kramer}, {Moran}, {Heckman},
  {Schiminovich}, {Schuster}, \& {Wuyts}}]{2012ApJ...758...73S}
{Saintonge}, A., {Tacconi}, L.~J., {Fabello}, S., {et~al.} 2012, \apj, 758, 73

\bibitem[{{Saintonge} {et~al.}(2013){Saintonge}, {Lutz}, {Genzel}, {Magnelli},
  {Nordon}, {Tacconi}, {Baker}, {Bandara}, {Berta}, {F{\"o}rster Schreiber},
  {Poglitsch}, {Sturm}, {Wuyts}, \& {Wuyts}}]{2013ApJ...778....2S}
{Saintonge}, A., {Lutz}, D., {Genzel}, R., {et~al.} 2013, \apj, 778, 2

\bibitem[{{Saintonge} {et~al.}(2016){Saintonge}, {Catinella}, {Cortese},
  {Genzel}, {Giovanelli}, {Haynes}, {Janowiecki}, {Kramer}, {Lutz},
  {Schiminovich}, {Tacconi}, {Wuyts}, \& {Accurso}}]{2016MNRAS.462.1749S}
{Saintonge}, A., {Catinella}, B., {Cortese}, L., {et~al.} 2016, \mnras, 462,
  1749

\bibitem[{Saintonge {et~al.}(2017)Saintonge, Catinella, Tacconi, Kauffmann,
  Genzel, Cortese, Dav{\'e}, Fletcher, Graci{\'a}-Carpio, Kramer, Heckman,
  Janowiecki, Lutz, Rosario, Schiminovich, Schuster, Wang, Wuyts, Borthakur,
  Lamperti, \& Roberts-Borsani}]{Saintonge:2017iz}
Saintonge, A., Catinella, B., Tacconi, L.~J., {et~al.} 2017, The Astrophysical
  Journal Supplement Series, 233, 0

\bibitem[{{Salim} {et~al.}(2007){Salim}, {Rich}, {Charlot}, {Brinchmann},
  {Johnson}, {Schiminovich}, {Seibert}, {Mallery}, {Heckman}, {Forster},
  {Friedman}, {Martin}, {Morrissey}, {Neff}, {Small}, {Wyder}, {Bianchi},
  {Donas}, {Lee}, {Madore}, {Milliard}, {Szalay}, {Welsh}, \&
  {Yi}}]{2007ApJS..173..267S}
{Salim}, S., {Rich}, R.~M., {Charlot}, S., {et~al.} 2007, \apjs, 173, 267

\bibitem[{{Salim} {et~al.}(2016){Salim}, {Lee}, {Janowiecki}, {da Cunha},
  {Dickinson}, {Boquien}, {Burgarella}, {Salzer}, \&
  {Charlot}}]{2016ApJS..227....2S}
{Salim}, S., {Lee}, J.~C., {Janowiecki}, S., {et~al.} 2016, \apjs, 227, 2

\bibitem[{{S{\'a}nchez}(2020)}]{2020ARA&A..5812120S}
{S{\'a}nchez}, S.~F. 2020, \araa, 58, annurev

\bibitem[{{Santini} {et~al.}(2014){Santini}, {Maiolino}, {Magnelli}, {Lutz},
  {Lamastra}, {Li Causi}, {Eales}, {Andreani}, {Berta}, {Buat}, {Cooray},
  {Cresci}, {Daddi}, {Farrah}, {Fontana}, {Franceschini}, {Genzel}, {Granato},
  {Grazian}, {Le Floc'h}, {Magdis}, {Magliocchetti}, {Mannucci}, {Menci},
  {Nordon}, {Oliver}, {Popesso}, {Pozzi}, {Riguccini}, {Rodighiero}, {Rosario},
  {Salvato}, {Scott}, {Silva}, {Tacconi}, {Viero}, {Wang}, {Wuyts}, \&
  {Xu}}]{2014A&A...562A..30S}
{Santini}, P., {Maiolino}, R., {Magnelli}, B., {et~al.} 2014, \aap, 562, A30

\bibitem[{{Sargent} {et~al.}(2014){Sargent}, {Daddi}, {B{\'e}thermin},
  {Aussel}, {Magdis}, {Hwang}, {Juneau}, {Elbaz}, \& {da
  Cunha}}]{2014ApJ...793...19S}
{Sargent}, M.~T., {Daddi}, E., {B{\'e}thermin}, M., {et~al.} 2014, \apj, 793,
  19


\bibitem[{{Schiminovich} {et~al.}(2007){Schiminovich}, {Wyder}, {Martin},
  {Johnson}, {Salim}, {Seibert}, {Treyer}, {Budav{\'a}ri}, {Hoopes},
  {Zamojski}, {Barlow}, {Forster}, {Friedman}, {Morrissey}, {Neff}, {Small},
  {Bianchi}, {Donas}, {Heckman}, {Lee}, {Madore}, {Milliard}, {Rich}, {Szalay},
  {Welsh}, \& {Yi}}]{2007ApJS..173..315S}
{Schiminovich}, D., {Wyder}, T.~K., {Martin}, D.~C., {et~al.} 2007, \apjs, 173,
  315

\bibitem[{{Schinnerer} {et~al.}(2016){Schinnerer}, {Groves}, {Sargent},
  {Karim}, {Oesch}, {Magnelli}, {LeFevre}, {Tasca}, {Civano}, {Cassata}, \&
  {Smol{\v{c}}i{\'c}}}]{2016ApJ...833..112S}
{Schinnerer}, E., {Groves}, B., {Sargent}, M.~T., {et~al.} 2016, \apj, 833, 112

\bibitem[{{Schmidt}(1959)}]{1959ApJ...129..243S}
{Schmidt}, M. 1959, \apj, 129, 243

\bibitem[{{Schreiber} {et~al.}(2015){Schreiber}, {Pannella}, {Elbaz},
  {B{\'e}thermin}, {Inami}, {Dickinson}, {Magnelli}, {Wang}, {Aussel}, {Daddi},
  {Juneau}, {Shu}, {Sargent}, {Buat}, {Faber}, {Ferguson}, {Giavalisco},
  {Koekemoer}, {Magdis}, {Morrison}, {Papovich}, {Santini}, \&
  {Scott}}]{2015A&A...575A..74S}
{Schreiber}, C., {Pannella}, M., {Elbaz}, D., {et~al.} 2015, \aap, 575, A74

\bibitem[{{Schruba} {et~al.}(2011){Schruba}, {Leroy}, {Walter}, {Bigiel},
  {Brinks}, {de Blok}, {Dumas}, {Kramer}, {Rosolowsky}, {Sand strom},
  {Schuster}, {Usero}, {Weiss}, \& {Wiesemeyer}}]{2011AJ....142...37S}
{Schruba}, A., {Leroy}, A.~K., {Walter}, F., {et~al.} 2011, \aj, 142, 37

\bibitem[{{Scoville} {et~al.}(2016){Scoville}, {Sheth}, {Aussel}, {Vanden
  Bout}, {Capak}, {Bongiorno}, {Casey}, {Murchikova}, {Koda},
  {{\'A}lvarez-M{\'a}rquez}, {Lee}, {Laigle}, {McCracken}, {Ilbert}, {Pope},
  {Sanders}, {Chu}, {Toft}, {Ivison}, \& {Manohar}}]{2016ApJ...820...83S}
{Scoville}, N., {Sheth}, K., {Aussel}, H., {et~al.} 2016, \apj, 820, 83

\bibitem[{{Scoville} {et~al.}(2017){Scoville}, {Lee}, {Vanden Bout},
  {Diaz-Santos}, {Sanders}, {Darvish}, {Bongiorno}, {Casey}, {Murchikova},
  {Koda}, {Capak}, {Vlahakis}, {Ilbert}, {Sheth}, {Morokuma-Matsui}, {Ivison},
  {Aussel}, {Laigle}, {McCracken}, {Armus}, {Pope}, {Toft}, \&
  {Masters}}]{2017ApJ...837..150S}
{Scoville}, N., {Lee}, N., {Vanden Bout}, P., {et~al.} 2017, \apj, 837, 150

\bibitem[{{Shi} {et~al.}(2011){Shi}, {Helou}, {Yan}, {Armus}, {Wu}, {Papovich},
  \& {Stierwalt}}]{2011ApJ...733...87S}
{Shi}, Y., {Helou}, G., {Yan}, L., {et~al.} 2011, \apj, 733, 87

\bibitem[{{Shi} {et~al.}(2018){Shi}, {Yan}, {Armus}, {Gu}, {Helou}, {Qiu},
  {Gwyn}, {Stierwalt}, {Fang}, {Chen}, {Zhou}, {Wu}, {Zheng}, {Zhang}, {Gao},
  \& {Wang}}]{2018ApJ...853..149S}
{Shi}, Y., {Yan}, L., {Armus}, L., {et~al.} 2018, \apj, 853, 149

\bibitem[{{Silk}(1997)}]{1997ApJ...481..703S}
{Silk}, J. 1997, \apj, 481, 703

\bibitem[{{Solomon} \& {Sage}(1988)}]{1988ApJ...334..613S}
{Solomon}, P.~M., \& {Sage}, L.~J. 1988, \apj, 334, 613

\bibitem[{{Speagle} {et~al.}(2014){Speagle}, {Steinhardt}, {Capak}, \&
  {Silverman}}]{2014ApJS..214...15S}
{Speagle}, J.~S., {Steinhardt}, C.~L., {Capak}, P.~L., \& {Silverman}, J.~D.
  2014, \apjs, 214, 15

\bibitem[{{Tacconi} {et~al.}(2020){Tacconi}, {Genzel}, \&
  {Sternberg}}]{2020arXiv200306245T}
{Tacconi}, L.~J., {Genzel}, R., \& {Sternberg}, A. 2020, arXiv e-prints,
  arXiv:2003.06245

\bibitem[{{Tacconi} {et~al.}(2010){Tacconi}, {Genzel}, {Neri}, {Cox}, {Cooper},
  {Shapiro}, {Bolatto}, {Bouch{\'e}}, {Bournaud}, {Burkert}, {Combes},
  {Comerford}, {Davis}, {Schreiber}, {Garcia-Burillo}, {Gracia-Carpio}, {Lutz},
  {Naab}, {Omont}, {Shapley}, {Sternberg}, \& {Weiner}}]{2010Natur.463..781T}
{Tacconi}, L.~J., {Genzel}, R., {Neri}, R., {et~al.} 2010, \nat, 463, 781

\bibitem[{{Tacconi} {et~al.}(2013){Tacconi}, {Neri}, {Genzel}, {Combes},
  {Bolatto}, {Cooper}, {Wuyts}, {Bournaud}, {Burkert}, {Comerford}, {Cox},
  {Davis}, {F{\"o}rster Schreiber}, {Garc{\'\i}a-Burillo}, {Gracia-Carpio},
  {Lutz}, {Naab}, {Newman}, {Omont}, {Saintonge}, {Shapiro Griffin}, {Shapley},
  {Sternberg}, \& {Weiner}}]{2013ApJ...768...74T}
{Tacconi}, L.~J., {Neri}, R., {Genzel}, R., {et~al.} 2013, \apj, 768, 74

\bibitem[{{Tacconi} {et~al.}(2018){Tacconi}, {Genzel}, {Saintonge}, {Combes},
  {Garc{\'\i}a-Burillo}, {Neri}, {Bolatto}, {Contini}, {F{\"o}rster Schreiber},
  {Lilly}, {Lutz}, {Wuyts}, {Accurso}, {Boissier}, {Boone}, {Bouch{\'e}},
  {Bournaud}, {Burkert}, {Carollo}, {Cooper}, {Cox}, {Feruglio}, {Freundlich},
  {Herrera-Camus}, {Juneau}, {Lippa}, {Naab}, {Renzini}, {Salome}, {Sternberg},
  {Tadaki}, {{\"U}bler}, {Walter}, {Weiner}, \& {Weiss}}]{2018ApJ...853..179T}
{Tacconi}, L.~J., {Genzel}, R., {Saintonge}, A., {et~al.} 2018, \apj, 853, 179

\bibitem[{Trussler {et~al.}(2019)Trussler, Maiolino, Maraston, Peng, Thomas,
  Goddard, \& Lian}]{Trussler_2019}
Trussler, J., Maiolino, R., Maraston, C., {et~al.} 2019, Monthly Notices of the
  Royal Astronomical Society, 491, 5406–5434

\bibitem[{{Ventou} {et~al.}(2017){Ventou}, {Contini}, {Bouch{\'e}}, {Epinat},
  {Brinchmann}, {Bacon}, {Inami}, {Lam}, {Drake}, {Garel}, {Michel-Dansac},
  {Pello}, {Steinmetz}, {Weilbacher}, {Wisotzki}, \&
  {Carollo}}]{2017A&A...608A...9V}
{Ventou}, E., {Contini}, T., {Bouch{\'e}}, N., {et~al.} 2017, \aap, 608, A9

\bibitem[{{Wang} {et~al.}(2011){Wang}, {Kauffmann}, {Overzier}, {Catinella},
  {Schiminovich}, {Heckman}, {Moran}, {Haynes}, {Giovanelli}, \&
  {Kong}}]{2011MNRAS.412.1081W}
{Wang}, J., {Kauffmann}, G., {Overzier}, R., {et~al.} 2011, \mnras, 412, 1081

\bibitem[{{Whitaker} {et~al.}(2012){Whitaker}, {van Dokkum}, {Brammer}, \&
  {Franx}}]{2012ApJ...754L..29W}
{Whitaker}, K.~E., {van Dokkum}, P.~G., {Brammer}, G., \& {Franx}, M. 2012,
  \apjl, 754, L29

\bibitem[{{Whitaker} {et~al.}(2014){Whitaker}, {Franx}, {Leja}, {van Dokkum},
  {Henry}, {Skelton}, {Fumagalli}, {Momcheva}, {Brammer}, {Labb{\'e}},
  {Nelson}, \& {Rigby}}]{2014ApJ...795..104W}
{Whitaker}, K.~E., {Franx}, M., {Leja}, J., {et~al.} 2014, \apj, 795, 104

\bibitem[{{White}(1984)}]{1984ApJ...286...38W}
{White}, S.~D.~M. 1984, \apj, 286, 38

\bibitem[{{Wiklind} {et~al.}(2019){Wiklind}, {Ferguson}, {Guo}, {Koo},
  {Kocevski}, {Mobasher}, {Brammer}, {Kassin}, {Koekemoer}, {Giavalisco},
  {Papovich}, {Ravindranath}, {Faber}, {Freundlich}, \& {de
  Mello}}]{2019ApJ...878...83W}
{Wiklind}, T., {Ferguson}, H.~C., {Guo}, Y., {et~al.} 2019, \apj, 878, 83

\bibitem[{{Wong} \& {Blitz}(2002)}]{2002ApJ...569..157W}
{Wong}, T., \& {Blitz}, L. 2002, \apj, 569, 157

\bibitem[{{Wyder} {et~al.}(2009){Wyder}, {Martin}, {Barlow}, {Foster},
  {Friedman}, {Morrissey}, {Neff}, {Neill}, {Schiminovich}, {Seibert},
  {Bianchi}, {Donas}, {Heckman}, {Lee}, {Madore}, {Milliard}, {Rich}, {Szalay},
  \& {Yi}}]{2009ApJ...696.1834W}
{Wyder}, T.~K., {Martin}, D.~C., {Barlow}, T.~A., {et~al.} 2009, \apj, 696,
  1834

\bibitem[{{Yesuf} \& {Ho}(2019)}]{2019ApJ...884..177Y}
{Yesuf}, H.~M., \& {Ho}, L.~C. 2019, \apj, 884, 177

\bibitem[{{Zhang} {et~al.}(2019){Zhang}, {Peng}, {Ho}, {Maiolino}, {Dekel},
  {Guo}, {Mannucci}, {Li}, {Yuan}, {Renzini}, {Dou}, {Guo}, {Man}, \&
  {Li}}]{2019ApJ...884L..52Z}
{Zhang}, C., {Peng}, Y., {Ho}, L.~C., {et~al.} 2019, \apjl, 884, L52

\end{thebibliography}
\end{document}